\documentclass[a4paper, amsfonts, amssymb, amsmath, reprint, showkeys, nofootinbib, singlepage, prx, superscriptaddress]{revtex4-2}
\usepackage{braket}
\usepackage[english]{babel}
\usepackage[utf8]{inputenc}
\usepackage[colorinlistoftodos, color=green!40, prependcaption]{todonotes}
\usepackage[pdftex, pdftitle={Article}, pdfauthor={Author}]{hyperref} 
\usepackage{mathrsfs}
\usepackage{bm}
\usepackage{nicefrac}
\usepackage{subcaption}
\newcommand{\MarkOrd}{\mathscr{M}}
\DeclareMathOperator{\Var}{Var}
\DeclareMathOperator{\Tr}{Tr}
\newcommand{\stilde}[1]{\Tilde{s}_{#1}}
\newcommand{\up}{\uparrow}
\newcommand{\down}{\downarrow}

\DeclareMathOperator{\Cov}{Cov}

\begin{document}
\title{Fisher information of correlated stochastic processes}

\author{Marco Radaelli}
    \email[]{radaellm@tcd.ie}
    \affiliation{School of Physics, Trinity College Dublin, Dublin 2, Ireland}
\author{Gabriel T. Landi}
    \email[]{gtlandi@gmail.com}
    \affiliation{School of Physics, Trinity College Dublin, Dublin 2, Ireland}
    \affiliation{Instituto de Física da Universidade de São Paulo, 05314-970 São Paulo, Brazil}
    \affiliation{Department of Physics and Astronomy, University of Rochester, Rochester, New York 14627, USA}
\author{Kavan Modi}
    \affiliation{School of Physics \& Astronomy, Monash University, Clayton, VIC 3800, Australia}
    \affiliation{Centre for Quantum Technology, Transport for New South Wales, Sydney, NSW 2000, Australia}
\author{Felix Binder}
    \email[]{quantum@felix-binder.net}
    \affiliation{School of Physics, Trinity College Dublin, Dublin 2, Ireland}

\begin{abstract}
    Many real-world tasks include some kind of parameter estimation, i.e., the determination of a parameter encoded in a probability distribution. Often, such probability distributions arise from stochastic processes. For a stationary stochastic process with temporal correlations, the random variables that constitute it are identically distributed but not independent. This is the case, for instance, for quantum continuous measurements. 
    In this article, we derive the asymptotic Fisher information rate for a stationary process with finite Markov order. We give a precise expression for this rate which is determined by the process' conditional distribution up to its Markov order. Second, we demonstrate with suitable examples that correlations may both enhance or hamper the metrological precision. Indeed, unlike for entropic information quantities, in general nothing can be said about the sub- or super-additivity of the joint Fisher information in the presence of correlations. To illustrate our results, we apply them to thermometry on an Ising spin chain, considering nearest-neighbour and next-to-nearest neighbour coupling. In this case, the asymptotic Fisher information rate is directly connected to the specific heat capacity of the spin chain. We observe that the presence of correlations strongly enhances the estimation precision in an anti-ferromagnetic chain, while in a ferromagnetic chain this is not the case.
\end{abstract}

\keywords{information theory, metrology, Fisher information, spin chains}

\maketitle

\section*{Introduction}
In many experimentally meaningful scenarios, a quantity of interest cannot be accessed directly via a measurement, but only retrieved from a sample of data; this problem is commonly referred to as \textit{estimation}. Examples arise in many disciplines: for instance, one might be interested in estimating financial parameters from the time series of stock prices, the lethality of epidemic diseases from real-world hospital data, or the Rabi frequency of a two level atom. Estimation theory \cite{Kay1993,VanTrees2004,Cover1999, Demkowicz2020} provides safe formal grounds for this estimation procedure, and gives ways to compute the maximum achievable precision that can be attained for a certain parameter, given the probability distribution from which the sample is extracted. The variance $\sigma_\theta^2$ of an unbiased estimator (inverse to the precision) for the parameter $\theta$ is subject to the Cramér-Rao bound~\cite{Fisher1922, Cramer1946},
\begin{equation}\label{CRB_basic}
    \sigma_\theta^2 \geqslant \frac{1}{F_{1:N}(\theta)},
\end{equation}
where $F_{1:N}(\theta)$ is the Fisher information associated to $N$ outcomes. In the case of i.i.d. random variables, the Cramér-Rao bound simply reduces to its most well-known version, given by $F_{1:N}(\theta) = N F_1(\theta)$. However, this is not true when, as often happens in the real world, processes exhibit correlations, where past values influence the future behaviour, possibly up to a fixed number of steps. 

Being able to quantify how many past outcomes have to be recorded to retrieve an accurate description of the future statistics is of crucial importance for data compression purposes, but also as a way to understand the structure and complexity of the underlying model~\cite{Crutchfield1989,Shalizi2001,Crutchfield2012}. In the context of discrete-time classical stochastic processes, the notion of memory length is expressed by the Markov order \cite{James2014}.

An important scenario featuring such correlated outcomes is that of quantum systems subject to continuous weak measurements~\cite{Jacobs_2014,Wiseman_2009}. 
This is common place in various experimental platforms, from quantum optics~\cite{PhysRevLett.108.243602,PhysRevA.96.063402,PhysRevA.97.013408} to mesoscopic conductors~\cite{Goan_2001}. It involves a stream of outcomes whose correlations crucially depend on the underlying quantum properties of the model. 
A famous example is the emission of light from quantum dots, which is known to present intermittencies~\cite{Nirmal_1996} (called \emph{blinking}) with very long range correlations~\cite{Yuan_2018,Munoz_2021}.
Several papers have recently analyzed the metrological properties of continuous measurements~\cite{Gammelmark_2013,Gammelmark2014,Kiilerich_2014,Kiilerich_2016,Burgarth_2015,Godley_2022,Ilias_2022,Seah_2019,Strasberg2020thermodynamicsof}, aiming both at constructing useful estimation strategies, as well as establishing the ultimate bounds in precision. 

The application of estimation theory to correlated processes is still  widely unexplored, and a general framework is lacking, except for very specific parameter encoding, such as location parameters~\cite{Carlen1991, Kagan1997}. In this work, we  specifically address this issue, and prove through suitable examples that the joint Fisher information is not in general sub-additive with respect to the marginal Fisher information; this fact sets the Fisher information apart from entropy-like information measures, for which sub-additivity is a fundamental requirement. 

The main result of our work, summarized in Eq.~\eqref{decomposition_Fisher_Markov}, is a general decomposition for the Fisher information for a stationary stochastic process of finite  Markov order $\MarkOrd$. 
This result showcases the interplay between the number of outcomes $N$, and the Markov order $\MarkOrd$. 
As we discuss in detail, this leads to  two fundamental consequences.
First, it shows that when $N\gg \MarkOrd$, the Fisher information will asymptotically always scale as 
\begin{equation}\label{main_result}
    F_{N}(\theta) \propto N f(\theta),
\end{equation}
where $f(\theta)$ is the Fisher information rate, which is independent of $N$. We here report the precise form of $f(\theta)$, its conditional dependence on preceding samples, and the non-asymptotic residual. Second, Eq.~\eqref{main_result} allows us to establish the sub- or super-additivity of $F_{1:N}$, by comparing $f$ with the Fisher information $F_1$ of a single outcome.
In fact, we show below by means of examples that in general $f \lessgtr F_1$; that is, unlike entropic information quantities, the Fisher information does not satisfy sub-additivity. Correlations can therefore be both advantageous or deleterious. The impact of these correlations in concrete estimation scenarios will be discussed in detail below.

This article is organized as follows. In Section \ref{sect:estimation_theory}, we review the basis of estimation theory, introduce the notion of unbiased estimator and the Fisher information. In Section \ref{sect:Fisher_in_presence_of_correlations} we discuss the behaviour of the Fisher information in the presence of correlations, showing with suitable toy models that in general it can exhibit super-additive or sub-additive behaviour with respect to the individual Fisher information. In Section \ref{sect:Markov_order} we introduce the notion of Markov order, and present our main result: a decomposition of the joint Fisher information in a finite-Markov order stochastic process. We discuss the consequences of the decomposition for the Cramér-Rao bound. To provide more intuition for our results, in Section \ref{sect:effects_correlations_sign}, we discuss how the sign of the correlations affects the behaviour of the Fisher information, and we show that, for the specific class of estimators based on the sample mean, anti-correlated processes usually yield higher Fisher information. Finally, in Section \ref{sect:application_thermometry_classical_spin_chains} we apply our results to the task of estimating temperature on an Ising spin chain, and show a direct connection between our decomposition of the Fisher information and the heat capacity per unit length (specific heat capacity) of the chain.

\section{Estimation Theory}
\label{sect:estimation_theory}
Consider the following problem. We are given a probability distribution $P_{\theta}(x_{1:N})$, describing the statistics of a sequence $X_{1:N} = X_1, \ldots, X_N$ (left- and right-inclusive notation will be assumed throughout this work) of $N$ random variables, and depending on a vector of real parameters $\theta=(\theta_1,\theta_2,\ldots,\theta_d)$. Our aim is to estimate, in the most precise  way possible, the vector $\theta$, by sampling from the probability distribution. 

In many scenarios, the parameters vector $\theta$ reduces to a scalar, such as in the case of thermometry we will deal with below. Other times, $\theta$ is a genuine vector (multi-parameter estimation), for example of Carthesian coordinates, velocities, rotation or torsion parameters. In the following, we will introduce the formalism of estimation theory in full generality in a multi-parameter fashion.

It is possible to show that the precision that can be achieved on the estimation of $\theta$ is upper-bounded by an intrinsic property of the probability distribution itself. An \textit{estimator} $T$ is a function which, given as an input a realization $x_{1:N}=\{x_1,x_2,\ldots,x_N\}$, outputs an estimate for the parameter vector:
\begin{equation}
    T: \{x_1,x_2,\ldots,x_N\} \mapsto \Tilde{\theta}.
\end{equation}
The Fisher information matrix is defined as \cite{Ly2017, Fisher1922, VanTrees2004} $F_{1:N}(\theta)$ as:
\begin{equation}
\label{Fisher_information_matrix}
\begin{split}
    \left[F_{1:N}(\theta)\right]_{ij} = &\\ \sum_{x_{1:N}} P_{\theta}(x_{1:N}) & \left(\frac{\partial }{\partial \theta_i} \log P_{\theta}(x_{1:N}) \frac{\partial}{\partial \theta_j}\log P_{\theta}(x_{1:N})\right).
\end{split}
\end{equation}
That is, the covariance between the two first derivatives (with respect to the parameter vector components $\theta_i$ and $\theta_j$) of the logarithms of the probability distribution (i.e., the \textit{surprisal}). 

The importance of the Fisher information is mainly due to its role in the expression of the Cramér-Rao bound, the ultimate limit to the precision of any estimate of the parameters. 
We define the \textit{bias} of an estimator \cite{Cover1999} as the average difference between the estimate of the parameter vector and the true value of the parameter vector itself:
\begin{equation}
    b(T) = \langle T(X_{1:N}) - \theta \rangle.
\end{equation}
An estimator is \textit{unbiased} if its bias is equal to zero. Given an unbiased estimator $T$, we can define its \textit{covariance matrix} $\sigma^2$ as the matrix \cite{VanTrees2004}
\begin{equation}
    \sigma_{ij}^2 = \sum_{x_{1:N}} P_{\theta}(x_{1:N}) \left(\left[T(x_{1:N})\right]_i - \theta_i\right)\left(\left[T(x_{1:N})\right]_j - \theta_j\right).
\end{equation}
The Cramér-Rao bound then reads~\cite{Cramer1946,Rao1973} 
\begin{equation}
    \sigma^2 \geq \frac{1}{F_{1:N}(\theta)},
    \label{eq:cramer-rao}
\end{equation}
where $\frac{1}{F_{1:N}(\theta)}$ is the inverse of the Fisher information matrix, and the inequality means that the difference $\sigma^2 - \frac{1}{F_{1:N}(\theta)}$ is a positive semi-definite matrix.
Positive definiteness implies, in particular, that the variances of each $\theta_k$ will satisfy  Eq.~\eqref{CRB_basic}, with $F$ replaced by the $(k,k)$-th component of the Fisher information matrix.

\begin{figure*}[htbp]
    \centering
    \includegraphics[trim=2cm 6.5cm 2cm 2cm, clip=true, width=\textwidth]{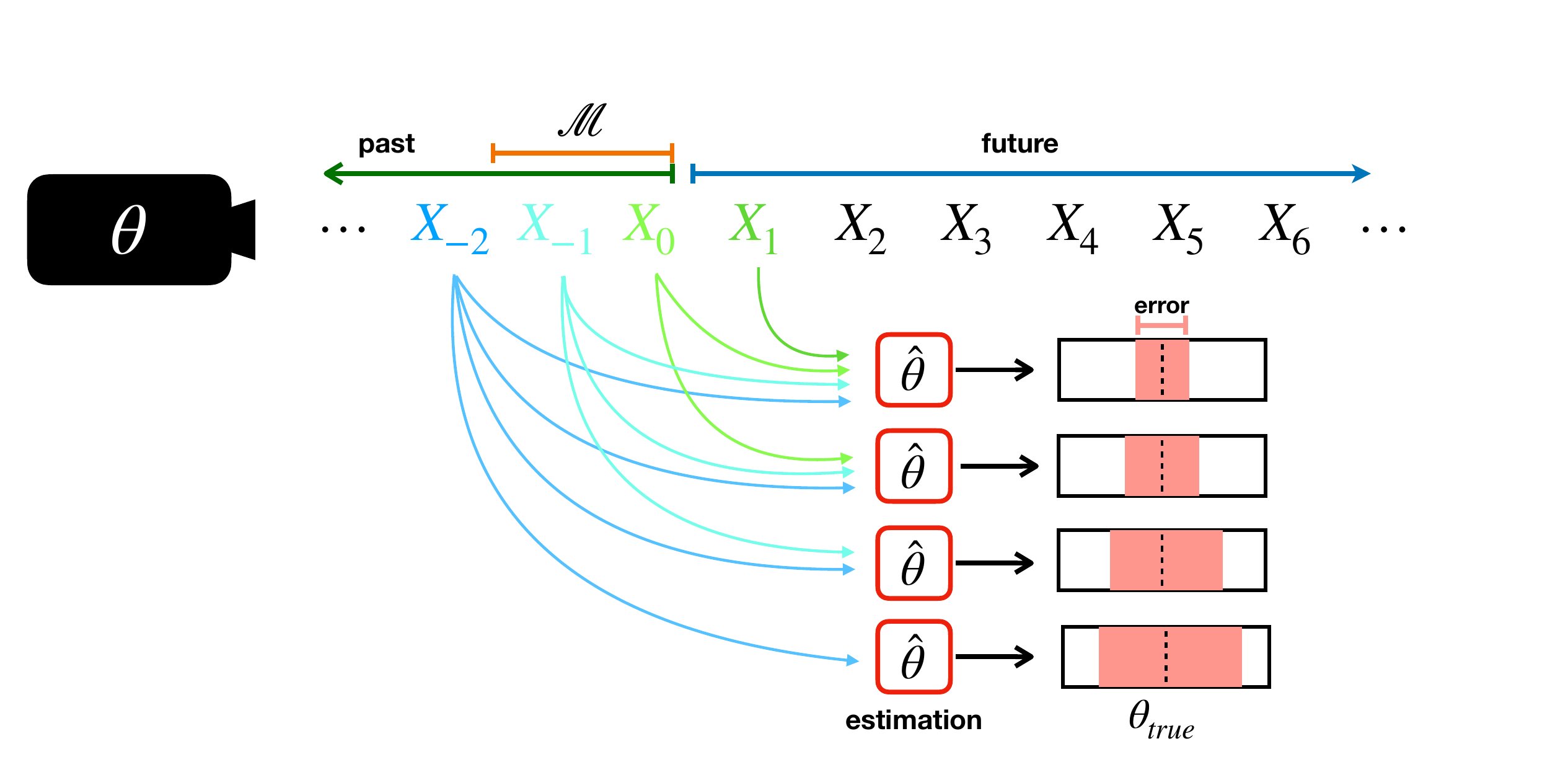}
    \caption{
    Sequential parameter estimation from correlated outcomes. The picture denotes a black box outputting outcomes $X_i$ that depend on some unknown parameter $\theta$. The outcomes, however, are not statistically independent, but instead have a finite Markov order $\MarkOrd$ (in this example $\MarkOrd=2$). 
    This encompass various scenarios in both classical and quantum processes. For example, it may represent the output current of a quantum continuous measurement. Or the outcomes of sequential local measurements in a quantum spin chain (Sec.~\ref{sect:application_thermometry_classical_spin_chains}). 
    The goal of this paper is to understand how  correlations in the outcomes affect the precision in estimating $\theta$.
    }
    \label{fig:scheme_sequential}
\end{figure*}

\section{Fisher information in  presence of correlations}
\label{sect:Fisher_in_presence_of_correlations}
The simplest parameter estimation scenario is that of independent and identically distributed (i.i.d.) random variables, for which factorization of the joint probability distribution
\begin{equation}
    \label{factorization_prob_distr}
    P_\theta(X_1,\ldots,X_N) = P_\theta(X_1)\ldots P_\theta(X_N),
\end{equation}
holds. In this case the Fisher information matrix behaves additively:
\begin{equation}
    F_{1:N}(\theta) = N F_1(\theta),
    \label{additivity_Fisher}
\end{equation}
as proven in Appendix \ref{sect:proof_decomposition_Fisher}, where $F_1(\theta)$ is the Fisher information matrix corresponding to the probability distribution of $P_\theta(X)$.
The Cramér-Rao bound (Eq.~\ref{eq:cramer-rao}) in this case expresses the  variance in $\theta$ scaling inversely with $N$:
\begin{align}
    \sigma_\theta^2\geq\frac{1}{NF_1(\theta)}
    \label{Cramér_Rao_iid}
\end{align}

In the presence of correlations
the factorization of the joint probability distribution no longer holds.
Instead, we define the conditional Fisher information matrix between two random variables $X_1$ and $X_2$,  as
\begin{equation}
\begin{split}
    & \left[F_{X_2\vert X_1}(\theta)\right]_{ij} = \sum_{x_1, x_2} P_\theta(x_1, x_2) \cdot \\
    & \cdot \left[\frac{\partial}{\partial \theta_i} \log P_\theta(x_2\vert x_1)\frac{\partial}{\partial\theta_j}\log P_\theta(x_2\vert x_1)\right].
\end{split}
\end{equation}
It is possible to show \cite{Micadei2015} that the joint and the conditional Fisher information matrices between two random variables are connected by the relation
\begin{equation}
  F_{X_2\vert X_1}(\theta)  =  F_{X_1,X_2}(\theta) - F_{X_1}(\theta),
\end{equation}
which may serve as an alternative definition of $F_{X_2\vert X_1}(\theta)$. The single-parameter case was treated in Ref.~\cite{Micadei2015}. A proof for the multi-parameter case can be found in Appendix \ref{sect:proof_decomposition_Fisher} (we are not aware of this general proof in the literature). 

It directly follows that the joint Fisher information matrix of $N$ outcomes can be decomposed as \cite{Zegers2015}
\begin{equation}
    F_{1:N}(\theta) = F_1(\theta) + \sum_{k=2}^{N} F_{k\vert 1:k-1}(\theta),
    \label{decomposition_Fisher}
\end{equation}
valid also in the case of multi-parametric estimation. If the variables are independent, this reduces back to Eq.~\eqref{additivity_Fisher}.

Until this point, our treatment has considered the case of multi-parameter estimation in full generality. From now on, for the sake of clarity we will restrict our attention to single-parameter estimation, and we will therefore think of the Fisher information as a scalar quantity. Our results can be trivially generalized to a multi-parameter scenario.

\subsection{Lack of sub-additivity for the Fisher information}

In general, nothing can be said  about the inequality
\begin{equation}
    F_{X_1 \ldots X_N}(\theta) \overset{?}{\lessgtr} F_{X_1}(\theta) + \ldots F_{X_N}(\theta).
\end{equation}
In this sense, the Fisher information does not behave as an entropy-like quantity, for which convexity is a fundamental requirement \cite{Cover1999}. To illustrate this idea, we consider a simple example based on Gaussian random variables $X,Y$, with identical mean $\mu$ and a covariance matrix $\sigma$ defined as
\begin{equation}
    \sigma = \gamma_0 \begin{pmatrix}1 & \rho \\ \rho & 1\end{pmatrix}
\end{equation}
for $\rho \in [-1, 1]$ and $\gamma_0 > 0$. 
The task at hand is then the estimation of the mean value $\mu$. The joint Fisher information for Gaussian random variables with mean vector $\vec{\mu}$ and covariance matrix $\sigma$ is well known, and given by $F = \left(\partial_\theta \vec{\mu}\right)^T \sigma^{-1} \left(\partial_\theta \vec{\mu}\right) + \frac{1}{2} {\rm tr}\big\{ \sigma^{-1} (\partial_\theta \sigma)  \sigma^{-1} (\partial_\theta \sigma) \big\}$~\cite{Kay1993}.
Applying this to the estimation of $\mu$ itself yields
\begin{equation}
    F_{X,Y}(\mu) = \frac{1}{\gamma_0} \frac{2}{1+\rho},
\end{equation}
The marginal Fisher information, on the other hand, is given by:
\begin{equation}
    F_{X}(\mu) = F_Y(\mu) = \frac{1}{\gamma_0}.
\end{equation}
Whence 
\begin{equation}\label{gaussian_two_variables_ratio}
    \frac{F_{X,Y}(\mu)}{F_{X}(\mu)+F_{Y}(\mu)} = \frac{1}{1+\rho}.
\end{equation}
It is therefore clear that we obtain a super-additive behaviour ($F_{X,Y} > F_X + F_Y$) when the variables are anti-correlated, $\rho<0$, and a sub-additive behaviour ($F_{X,Y} < F_X + F_Y$) when they are correlated, $\rho>0$. 
For instance, if $X$ and $Y$ are perfectly correlated ($\rho=1$), we get $F_{X,Y} = F_X$: We learn nothing from $Y$ that we hadn't already learned from $X$. 
The fact that we can get super-additive behavior for $\rho <0$, though, is not at all intuitive. We will revisit this in Sec.~\ref{sect:effects_correlations_sign}, where we  show that it happens because for anti-correlated variables the errors tend to cancel out, while in the correlated case they tend to add up.

\section{The Markov order}
\label{sect:Markov_order}
Let us consider a family of stochastic processes $\ldots,X_0,X_1,\ldots,X_N,\ldots = \overleftrightarrow{X}$, fully represented by the overall joint parameter-dependent probability distribution $P_\theta(\overleftrightarrow{x})$. Here, we idealize stochastic processes as bi-infinite. A process is \textit{stationary} \cite{Cover1999} if any marginal of the overall joint probability distribution is translationally invariant, i.e.:
\begin{equation}
P_\theta(X_{k:m}=x_{k:m}) = P_\theta(X_{k+l:m+l}=x_{k:m}),
\end{equation}
for any $k,m,l\in\mathbb{Z}, m\geq k$. 

Given a stationary stochastic process, one can define its Markov order \cite{James2014} as:
\begin{equation}
    \label{Markov_order}
    \MarkOrd = \min \left\{ \ell \mbox{ s.t. } P\left(X_{1}\vert X_{-\ell:0}\right)=P\left(X_{0}\vert X_{-\infty:0}\right) \right\},
\end{equation}
where $X_{-\infty:0}  = \overleftarrow{X}$ denotes the left-semi-infinite string of random variables in the stochastic process.
The concept is illustrated in Figure \ref{fig:scheme_sequential}. 
The definition of the Markov order implies the possibility to truncate any conditioning in the probabilities to a fixed number of steps in the past, representing the \textit{memory} of the system \cite{Taranto2018} (see also Eq.~\ref{eq:conditional_independence} below):
\begin{equation}
    P(\overrightarrow{X} \vert \overleftarrow{X}) = P(\overrightarrow{X} \vert X_{-\MarkOrd+1:0}),
\end{equation}
where $\overrightarrow{X} = X_{1:+\infty}$. For example, in a process of Markov order $\MarkOrd=2$, all knowledge about the future is encoded in the two `most recent' random variables in the past:
\begin{equation}
    P(\overrightarrow{X}\vert \overleftarrow{X}) = P(\overrightarrow{X} \vert X_{-1} X_0).
\end{equation}
Markov order $\MarkOrd=0$ represents independent random variables
\begin{equation}
    P(\overrightarrow{X}\vert \overleftarrow{X}) = P(\overrightarrow{X}). 
\end{equation}
Processes yielding Markov order $\MarkOrd \leq 1$ are customarily called \textit{Markov processes}. 

It is worth emphasizing that the Markov order only establishes conditional independence. Unconditionally, \emph{all} random variables in the chain are correlated (unless $\MarkOrd=0$).
These correlations, however, decay exponentially with their distance for all processes with finite Markov order.
For instance, in the case of Markov order $\MarkOrd = 1$ the process forms a Markov chain specified by the transition probabilities $Q_{xy} = P(X_{i+1}=x|X_i=y)$. 
For any two random variables $X_i$ and $X_j$, one will then have that (assuming $j>i$)
\begin{equation}
    P(X_j|X_i) = (Q^{j-i})_{X_j,X_i}. 
\end{equation}
Since $Q$ forms a stochastic matrix, its eigenvalues must have absolute value smaller than unity. 
Thus, the correlations between $X_i$ and $X_j$ will decay exponentially with $|j-i|$. 
The same reasoning also holds for processes having $\MarkOrd>1$, provided that $|j-i| > \MarkOrd$. 
 
The interplay between conditional and unconditional independence is subtle, and often counterintuitive. Some non-trivial examples are discussed in Appendix \ref{sect:Gaussian_Markov_process}. We also mention that, in general the Markov order is a \textit{structural} property of a stochastic process (or \textit{topological} in the language of computational mechanics~\cite{James2018}), independent of the precise choice of the parameters; other quantities like the (covariance-based) correlation length, on the contrary, usually exhibit a significant dependence on the parameters.

Alongside the probability-theoretic approach \eqref{Markov_order}, it is equivalently possible to define the Markov order from an information-theoretic perspective as \cite{Crutchfield2003}
\begin{equation}
    \label{entropy_theoretic_Markov_order}
    \MarkOrd =     \min\left\{\ell \mbox{ s.t. } H(X_{1:\ell}) = \ell h + E\right\}.
\end{equation}
Here, $H$ denotes the Shannon entropy of the distribution, while $h$ is the \textit{entropy rate}, defined as 
\begin{equation}
    \lim_{N\to\infty} \frac{1}{N} H(X_{1:N}).
\end{equation}
Finally, the \textit{excess entropy} $E$ is given by the mutual information between past and future
\begin{equation}
    E = \lim_{N\to\infty} I(X_{-N:0}; X_{1:N}).
\end{equation}
It follows that, for a process of finite Markov Order $\MarkOrd$, the memory renders past and future conditionally independent:
\begin{equation}
    I(\overrightarrow{X}; X_{-\infty:-\MarkOrd} \vert X_{-\MarkOrd+1:0}) = 0,
    \label{eq:conditional_independence}
\end{equation}
meaning that no information about the future is retained in the past beyond the Markov order. The relation between the information-theoretic and the probability-theoretic versions of the Markov order is discussed in Appendix \ref{sect:different_definitions_Markov_order}.

For a stationary process of finite Markov order $\MarkOrd$ we now consider the decomposition~\eqref{decomposition_Fisher} for the Fisher information for a string of $N\geq\MarkOrd$ outcomes $F_{1:N}$. In this case, all the conditional probabilities can be truncated at length $\MarkOrd$, and the same will happen to the conditional Fisher information. We are therefore left with:
\begin{equation}
    \begin{split}
        F_{1:N}(\theta) =& F_1(\theta) + \sum_{k=2}^N F_{k\vert 1:k-1}(\theta)  \\
         =& F_1(\theta) + F_{2\vert 1}(\theta) + F_{3\vert 1:2}(\theta) + \ldots \\
        & + F_{\MarkOrd\vert 1:\MarkOrd-1}(\theta) + \\ & + \sum_{k=1}^{N-\MarkOrd} F_{\MarkOrd+k\vert k-1:\MarkOrd+k}(\theta).
    \end{split}
\end{equation}
but, since the process is stationary, all the terms of the summation yield the same contribution, and, using \eqref{decomposition_Fisher}, we can directly write:
\begin{equation}
    F_{1:N}(\theta) = F_{1:\MarkOrd}(\theta) + (N-\MarkOrd) F_{\MarkOrd+1\vert 1:\MarkOrd}(\theta).
    \label{decomposition_Fisher_Markov}
\end{equation}
This is our main result. 
In the remainder of this section, we showcase its significance by discussing some of the main consequences. 

It is natural to write Eq.~\eqref{decomposition_Fisher_Markov} in the spirit of \eqref{entropy_theoretic_Markov_order}, as
\begin{equation}
    \label{rate_excess_Fisher}
    F_{1:N}(\theta) = N f(\theta) + \epsilon,
\end{equation}
defining the Fisher information rate $f(\theta)$ as
\begin{equation}
    f(\theta)=\lim_{N\to\infty} \frac{1}{N} F_{1:N}(\theta) = F_{\MarkOrd+1\vert 1:\MarkOrd}(\theta),
\end{equation}
and the excess Fisher information $\epsilon(\theta)$ as
\begin{equation}
    \epsilon(\theta) = F_{1:N}(\theta) - N f(\theta) = F_{1:\MarkOrd}(\theta) - \MarkOrd f(\theta).
\end{equation}
This expression leads to a clear analogy with \eqref{entropy_theoretic_Markov_order}; however, it is remarkable that, as already shown in the example of the two Gaussian random variables (in Sec.~\ref{sect:Fisher_in_presence_of_correlations}), $\epsilon$ can be both positive or negative, while the excess entropy $E$ is always positive.

To provide a concrete example, consider the case $\MarkOrd=1$. We then get 
\begin{align}
    f &= F_{2|1} = F_{12} - F_1 
    \\[0.2cm]
    \epsilon &= F_1 - f = 2F_1 - F_{12}. 
\end{align}
We must always have $F_{12} > F_1$, since extra information can only improve the estimation. 
However,  $\epsilon$ does not have a well defined sign because $F_{12} \lessgtr 2 F_1$ (or, what is equivalent $F_{2|1} \lessgtr F_1$). 

Exploiting the decomposition \eqref{decomposition_Fisher_Markov}, the Cramér-Rao bound \eqref{eq:cramer-rao} can be expressed as
\begin{equation}
\begin{split}
     &\sigma_{\theta}^2 \geq \frac{1}{F_{1:N}(\theta)}\\ &  = \frac{1}{F_{1:\MarkOrd}(\theta) + (N-\MarkOrd) F_{\MarkOrd+1\vert 1:\MarkOrd}(\theta)} \\ & \sim
     \frac{1}{N F_{\MarkOrd+1\vert 1:\MarkOrd}(\theta)}
\end{split}
\label{Cramer_Rao_Markov_Fisher}
\end{equation}
where the last approximation is valid in the large $N$ limit. Hence: for a finite-Markov order process, the maximum precision of the measurement, given by the Cramér-Rao bound, grows linearly with the number of measurements, in the limit of a large number of measurements. 

Since almost every stochastic process can always be assumed to have a finite (possibly large) Markov order, this means that for sufficiently large $N$ the precision will always eventually scale linearly with $N$. 
The remaining question, then, is how does $F_{\MarkOrd+1\vert 1:\MarkOrd}$ behave.
More concretely, we can aim to compare it with $F_1$, which would be the Fisher information that one would obtain if the outcomes were independent. 
Eq.~\eqref{gaussian_two_variables_ratio}, however, has already provided one illustration that nothing can in general be said about this.
If $F_{\MarkOrd+1\vert 1:\MarkOrd} > F_1$, then having correlations is advantageous for the estimation, while if $F_{\MarkOrd+1\vert 1:\MarkOrd}< F_1$ they are deleterious. 
We therefore now turn to additional examples, which aim to pinpoint which ingredients might lead to $F_{\MarkOrd+1\vert 1:\MarkOrd} \lessgtr F_1$.

\subsection{An example of  super- and sub-additive behaviours}

\begin{figure*}[h!tbp]
    \centering
    \includegraphics[width=\textwidth, ]{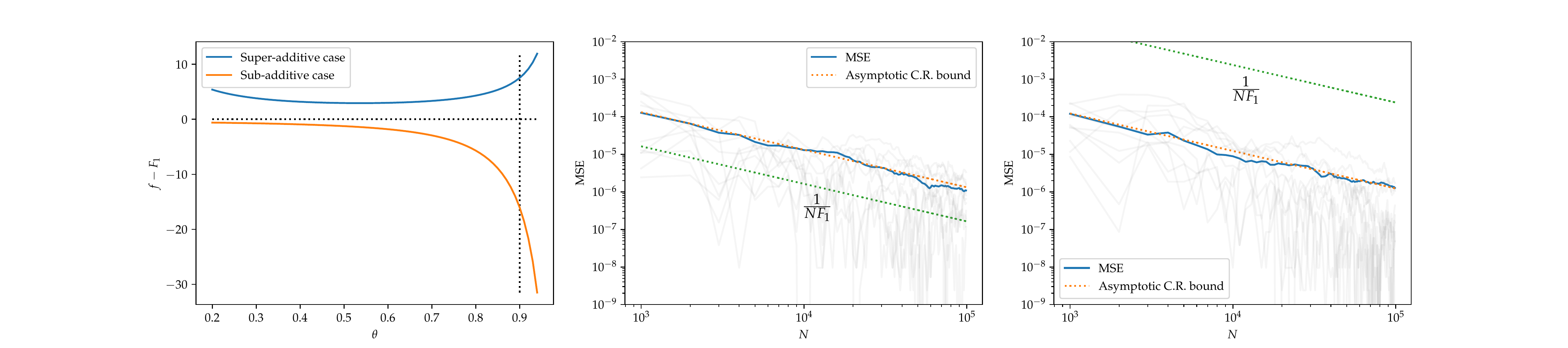}
    \caption{Two toy-model estimation tasks using the Maximum Likelihood Estimator show the convergence of the expectation value of the MSE, obtained by averaging over 50 iterations of the simulation, to the asymptotic behaviour given by \eqref{Cramer_Rao_Markov_Fisher}. The expression of the conditional probabilities yielding these processes are in the main text, and the leftt-most plot shows the corresponding sub- or super-additive behaviours of the Fisher information. The semi-transparent lines represent the MSE of single runs of the simulation. Notice that the bound is achieved only asymptotically, hence it is possible to have, for finite $N$, an MSE lower than the bound.}
    \label{fig:MSE_ToyModel_MLE}
\end{figure*}

We illustrate the above results with  minimal two-state models having Markov order $\MarkOrd=1$. For instance, a model having a sub-additive Fisher information is one given by the transition matrix:  
\begin{equation}
    P_\theta(X_{j+1}\vert X_j) = \begin{pmatrix}\theta & 1-\sqrt{\theta} \\ 1-\theta & \sqrt{\theta} \end{pmatrix}, \qquad \theta \in [0,1],
    \label{subadditive_example}
\end{equation}
Conversely, a model having a super-additive Fisher information is that given by 
\begin{equation}
    P_\theta(X_{j+1}\vert X_j) = \begin{pmatrix}\theta & 1-e^{-\theta/3} \\ 1-\theta & e^{-\theta/3} \end{pmatrix},\,\,\,\,\,\, \theta \in [0,1].
\end{equation}
The Fisher information sub- or super-additivity is spoiled by the difference $F_{1:2}-2F_1$, or equivalently $f - F_1$, represented for both the processes in the left-most plot of Figure \ref{fig:MSE_ToyModel_MLE} as a function of the parameter $\theta$.

The relevance of the sub- or super-additive behaviour becomes apparent if we consider an actual estimation task for these two models. 
It is well-known \cite{Cover1999, VanDerVaart2000} that the Maximum Likelihood Estimator always asymptotically attains the Cramér-Rao precision bound (also in the case of correlated variables, under adequate regularity conditions, see Appendix \ref{sect:MLE}); in Figure \ref{fig:MSE_ToyModel_MLE} we compare the Mean Squared Error given by the MLE estimation on the models with our asymptotic bound \eqref{Cramer_Rao_Markov_Fisher}.

The plots show two relevant facts: first, the convergence of the MSE of the MLE estimations to the asymptotic Cramér-Rao bound \ref{Cramer_Rao_Markov_Fisher} is verified; second, such a bound can be both much larger and much smaller than the i.i.d. Cramér-Rao bound for the specific case: once more, the Fisher information behaves sometimes super- and sometimes sub-additively.

The existence of processes yielding a sub-additive Fisher information could raise the question whether it could be possible to \textit{beat the system} by employing some estimator which, by ignoring the correlations, can achieve the i.i.d. Cramér-Rao bound. It turns out that this is not the case, as discussed in more details in Appendix \ref{sect:free_lunch}: the Fisher information (and consequently the Cramér-Rao bound) is an intrinsic property of the process, and cannot be circumvented by any choice of an estimator.

\subsection{Influence of the sign of correlations in the sample mean}
\label{sect:effects_correlations_sign}

Eq.~\eqref{gaussian_two_variables_ratio} provided an example where the sign of the correlations $\rho \lessgtr 0$ determined whether correlations are advantageous or not for the estimation process.
When considering a general estimator, it unfortunately turns out that a direct link cannot be established; however, an interesting insight can be gained by restricting our attention to a specific subclass of estimators, namely those based on the sample mean. The employment of the sample mean usually represents the simplest choice when asked to estimate something based on a stochastic process.

Let us consider a $\theta$-dependent stationary stochastic process $X_1, \ldots, X_N$; its sample mean is given by
\begin{equation}\label{sample_mean}
    Y = \frac{X_1 + \ldots + X_N}{N}.
\end{equation}
Let $\mu = \mathbb{E}[X_i]$ be the local expectation value, and $\sigma^2 = \Var[X_i]$ the local variance. Due to the stationarity requirement, the covariance between two random variables $X_i$ and $X_j$ depends only on their distance, let $\Cov[X_i, X_j] = C_{i-j}$. 
We refer to $C_i>0$ as  \textit{positively correlated} and $C_i<0$ as \textit{negatively correlated}.

Clearly, $\mathbb{E}[Y] = \mu$.
The variance, on the other hand, reads
\begin{equation}
    \Var[Y] = \frac{1}{N}\sigma^2 + \frac{2}{N^2}\sum_{i=1}^{N-1} (N-i) C_i,
\end{equation}
where we used the relation
\begin{equation}
    \sum_{i,j>i} \Cov[X_i, X_j] = \sum_{i=0}^{N-1}\sum_{j=i+1}^{N-1} C_{i-j} = \sum_{i=1}^{N-1} (N-i) C_i.
\end{equation}
Under appropriate hypotheses (discussed in Appendix \ref{sect:CLT}), a generalization of the Central Limit Theorem for correlated variables guarantees asymptotic normality of the  distribution of $Y$, with average $\mu$ and variance $\nu$, given by
\begin{equation}
    \nu = \frac{1}{N}\left[\sigma^2 + 2 \sum_{i=1}^{N-1} C_i\right].
\end{equation}
Under this normality assumption, the Fisher information of the sample mean $Y$ will be:
\begin{equation}
    F_Y(\theta) = N \left[\frac{1}{v} \left(\frac{\partial}{\partial\theta}\mu\right)^2 + \frac{1}{2N}\frac{1}{v^2}\left(\frac{\partial}{\partial\theta}v\right)^2\right],
\end{equation}
where $v = N \nu$. Notice that the second term is negligible in the $N\to\infty$ limit, hence we finally get a Fisher information
\begin{equation}\label{F_Y_final}
    F_Y(\theta) \sim N \left(\frac{\partial\mu}{\partial\theta}\right)^2 \frac{1}{\sigma^2 + 2 \sum_{i=1}^{N-1} C_i}.
\end{equation}

This result, which also appears in Ref.~\cite{kurdzialek_2021} in the context of super-resolution imaging theory, clearly illustrates the role that the sign of $C_i$ has on the Fisher information. 
Positively correlated variables ($C_i > 0$) increase the denominator and hence decrease $F_Y$, while negatively correlated $C_i$ do the opposite. 
\emph{For estimation, it is thus advantageous to have negatively correlated outcomes.} 
To understand the intuition behind this, consider Eq.~\eqref{sample_mean} and  write $X_i = \mu + \delta X_i$, where $\delta X_i$ represent the fluctuations around the mean. Then 
\begin{equation}\label{sample_mean2}
    Y = \mu + \frac{\delta X_1 + \ldots + \delta X_N}{N}.
\end{equation}
If two variables $\delta X_i$ and $\delta X_j$ are negatively correlated, then an outcome $\delta X_i>0$ implies there is a tendency to observe $\delta X_j<0$. 
As a consequence, \emph{errors tend to cancel out} in Eq.~\eqref{sample_mean2}. Conversely, for positively correlated variables, the errors tend to add up. 

Since $F_Y$ is the Fisher information for a restricted set of estimators, it must necessarily lower bound the full Fisher information, 
\begin{equation}
    F_Y \leqslant F_{1:N}.
\end{equation}
Using \eqref{decomposition_Fisher_Markov} we have therefore that, in the $N\to\infty$ limit, $F_Y \leq N F_{\MarkOrd+1\vert 1:\MarkOrd}$. Hence, 
\begin{equation}
F_{\MarkOrd+1\vert 1:\MarkOrd}(\theta) \geq     \left(\frac{\partial\mu}{\partial\theta}\right)^2 \frac{1}{\sigma^2 + 2 \sum_{i=1}^{N-1} C_i}.
\end{equation}
This provides a fundamental bound for the asymptotic conditional Fisher information, clearly showcasing the effects of correlations.

\begin{figure*}[h!tbp]
    \centering
    \includegraphics[trim=2cm 0cm 2cm 0cm, clip=true, width=\textwidth]{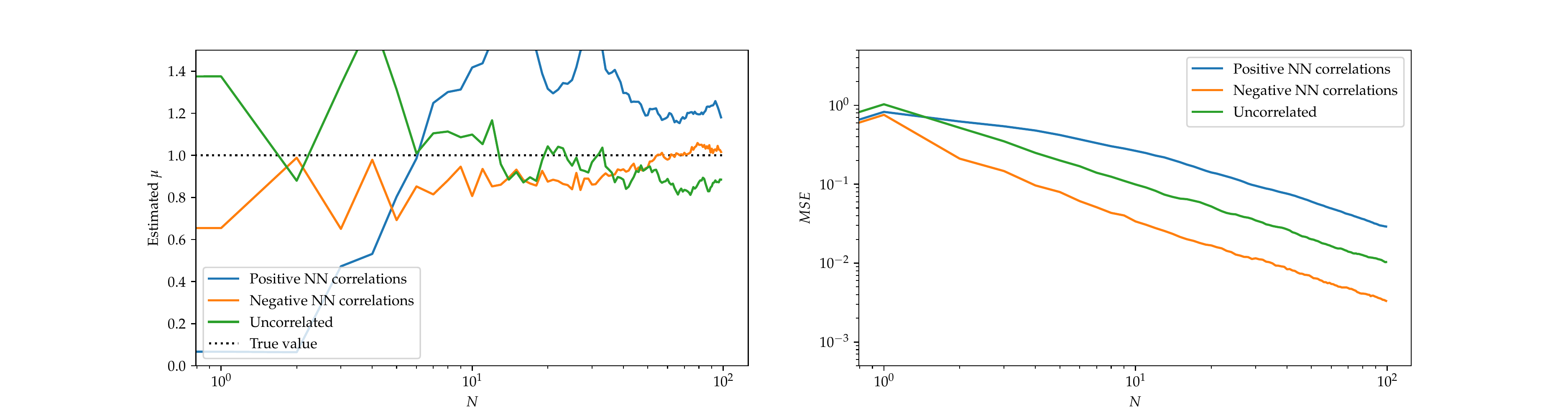}
    \caption{A simulated estimation task exploiting the sample mean on a Markov chain of Gaussian variables, for different values of the correlation parameter $\rho$. On the left, a single run of the estimation, on the right  the Mean Squared Error (MSE) is averaged over 1000 iterations of the task. For details about the construction of this Gaussian Markov process, see Appendix \ref{sect:Gaussian_Markov_process}.
    }
    \label{fig:MSE_Gaussian}
\end{figure*}

These ideas are illustrated in Figure \ref{fig:MSE_Gaussian}, which analyzes the estimation of the mean $\mu$ of a Gaussian process, with tunable nearest-neighbor correlations given by the parameter $\rho \in [0,1]$. Details on how to construct the process are discussed in Appendix \ref{sect:Gaussian_Markov_process}.
The left panel shows a single stochastic realization using the sample mean as an estimator. 
As can be seen, positively and negatively correlated samples seem to converge at a somewhat similar rate to the true parameter, while uncorrelated samples seem to converge more slowly. 
The impression from this single realization, however, is deceiving. 
In the right panel of Fig.~\ref{fig:MSE_Gaussian} we show the mean squared error (MSE) averaged over multiple realizations. It is now quite clear that negatively correlated samples yield much more precise estimates when compared to the uncorrelated case (notice the log scale), in agreement with the predictions of Eq.~\eqref{F_Y_final}. 
Conversely, positively correlated samples yield a clear and sizable disadvantage.

\section{Application to thermometry of spin chains}
\label{sect:application_thermometry_classical_spin_chains}

In this section, we apply the above detailed methods to the task of temperature estimation on an Ising spin chain~\cite{Ising1925}, consisting of $N$ 1/2-spins, with Hamiltonian 
\begin{equation}\label{hat_H}
    \hat{H} = \sum_j -B \hat{\sigma}_j^z - \sum_k J_k \sum_j \hat{\sigma}_j^z \hat{\sigma}_{j+k}^z,
\end{equation}
where $\sigma_j^z$ are Pauli matrices,
$B$ is the the external magnetic field, and $J_k$ is a $k$-distance coupling strength. 
The system is prepared in a thermal state 
\begin{equation}
    \hat{\rho} = \frac{1}{Z} e^{- \hat{H}/T},
\end{equation}
whose temperature $T$ we wish to estimate (i.e., it plays the role of the parameter $\theta$), and $Z$ is the partition function (we set the Boltzmann constant to $k_B = 1$). 

Due to the structure of the Gibbs state $\rho$,  the best strategy for thermometry is always to measure the system in the energy basis~\cite{Correa_2015,Hovhannisyan_2021}. 
Eq.~\eqref{hat_H} is already diagonal, and thus a projective energy measurement is tantamount to $N$ local measurements in the basis of $\sigma_j^z$. 
The outcome is a string $\bm{s} = (s_1,\ldots,s_N)$, with $s_j = \pm 1$,  which occurs with probability 
\begin{equation}
    \label{thermal_state_spin_chain}
    p(\bm{s}) = \frac{1}{Z} \exp\left[-\frac{1}{T} H(\bm{s})\right],
\end{equation}
where
\begin{equation}
    \label{Hamiltonian_spin_chain}
    H(\bm{s}) = \sum_{j} -B s_j - \sum_{k} J_k \sum_{j} s_{j}s_{j+k}.
\end{equation}
Because of the interactions, the outcomes of each site are not statistically independent. 
Hence $p(\bm{s})$ forms a Markov chain, precisely of the form described in the previous section. 

For intuition, one may imagine that the spins are measured sequentially, from left to right as in Fig.~\ref{fig:scheme_sequential}. 
In this way, subsequent measurements (each yielding the current state of the measured spin) will progressively form a stochastic process. 
To apply the definition of the Markov order \eqref{Markov_order} to the spin chain, one first has to be able to marginalize the chain's thermal state \eqref{thermal_state_spin_chain} in order to express probability distributions for configurations of finite number of spins. The procedure for doing so is based on the well-known transfer matrix method \cite{Kramers1941} and is detailed in~\cite{Feldman1998,Feldman1998b}. In Appendix \ref{sect:marginalisation_spin_chains}, we briefly summarize the main results. 

The problem of the Markov order of spin chains has been discussed in Ref.~\cite{James2014}. Let us first define the \textit{interaction range} $R$ of a spin chain as
\begin{equation}
    R = \min \left\{k \mbox{ s.t. } J_s = 0 \,\,\, \forall s > k\right\}.
\end{equation}
For instance, for nearest neighbor interactions $J_1 \neq 0$, while $J_s = 0$ for $s>1$, which would lead to $R = 1$. Naively, one might expect that $R=\MarkOrd$. However, this is not always true. In~\cite{Feldman1998} it was implicitly proved that $R\geq \MarkOrd$ for a completely general spin chain. Despite its intuitiveness, it is not clear whether the equality holds in all cases (as already noted in Ref.~\cite{Feldman1998b}). E.g. obvious exceptions appear for $T=0$ and $1/T=0$; however, these extremal, singular points bear no mathematical relevance in this context where $T$ is the parameter to be estimated. We leave a general proof of the full conditions under which $R=\MarkOrd$ as open question. A review of the discussion about the Markov order of classical spin chains can be found in Appendix~\ref{sect:Markov_order_spin_chains}.

Once we have determined the Markov order of a specific spin chain, for any number of consecutive spins larger than the Markov order the Fisher information scales according to \eqref{decomposition_Fisher_Markov}. 

If a system is in a thermal state, expressed by a Gibbs distribution \eqref{thermal_state_spin_chain}, then the well-known relation between the thermal Fisher information and the heat capacity $C$ holds~\cite{Correa_2015}:
\begin{equation}
    F(T) = \frac{C}{T^2},
\end{equation}
where $C$ is defined as:
\begin{equation}
    C = \frac{\partial \langle H \rangle}{\partial T}.
\end{equation}
It is important to remark that this relation holds exactly only if we take into account the whole chain, since the marginals of a Gibbs distribution are not, in general, of Gibbs form. 
However, the Fisher information decomposition \eqref{decomposition_Fisher_Markov} allows us to relate directly the heat capacity per unit length (i.e., the \textit{specific heat capacity}) to the Fisher information of suitable marginals.

Let us recall the decomposition of the Fisher information (Eq.~\eqref{rate_excess_Fisher}); defining the specific heat capacity as
\begin{equation}
    c = \frac{C(N)}{N},
\end{equation}
where $C(N)$ denotes the heat capacity for a chain of length $N$, we have:
\begin{equation}
    \frac{c}{T^2} = \frac{F_{1:N}(T)}{N} = \frac{1}{N}\left(N f + \epsilon\right).
\end{equation}
Taking the $N\to\infty$ limit, since neither $f$ nor $\epsilon$ depend on $N$, we then obtain:
\begin{equation}
    \lim_{N\to\infty} \frac{c}{T^2} = f = F_{\MarkOrd+1\vert 1:\MarkOrd}(T).
\end{equation}

\begin{figure}[htbp]
    \centering
    \includegraphics[trim=0.8cm 0cm 2cm 0cm, width=\columnwidth]{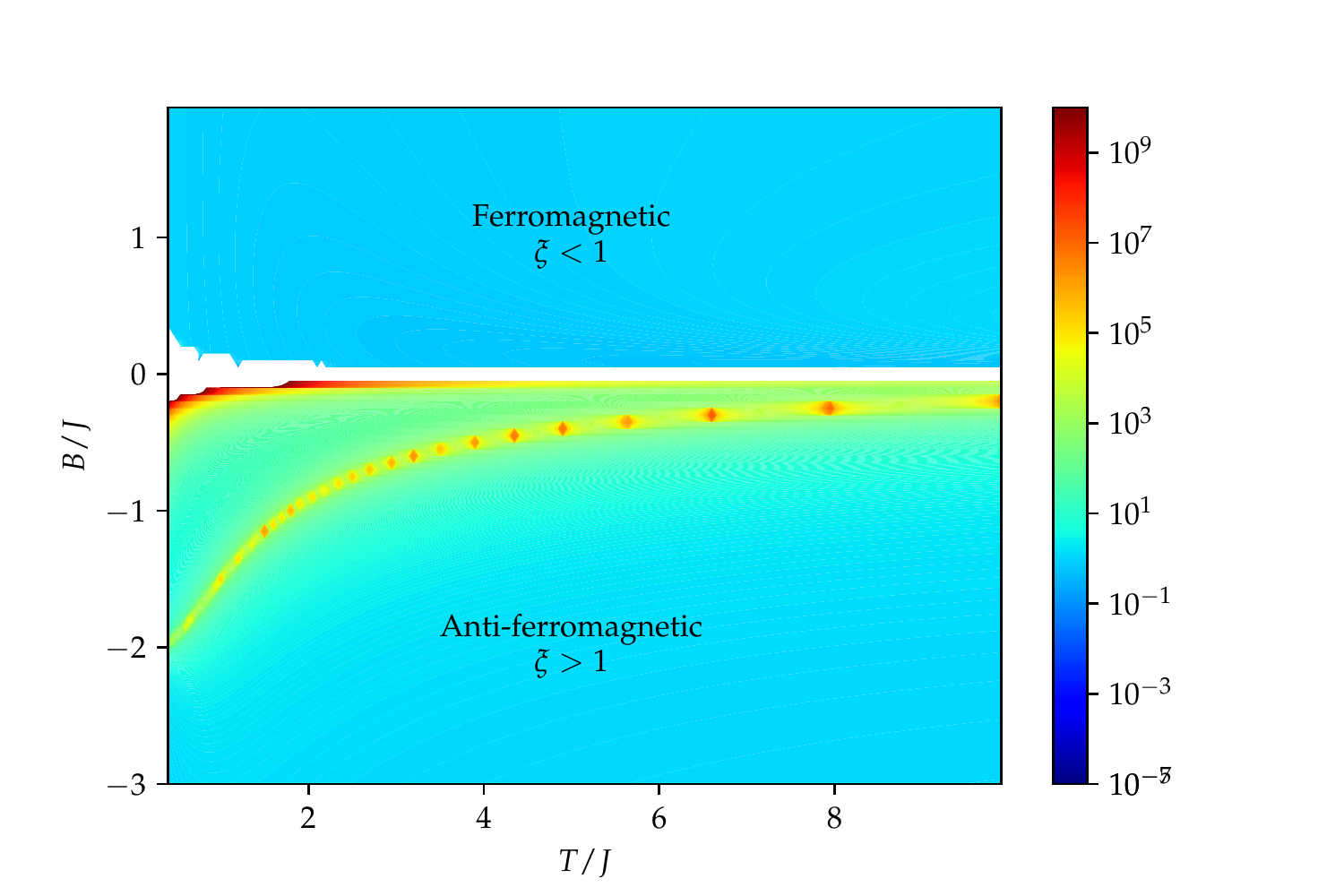}
    \caption{Logarithmic scale heatmap representing the value of the $\xi$-ratio for the additivity of the Fisher information on a nearest-neighbors Ising chain. Notice the difference between the ferromagnetic and the anti-ferromagnetic regimes. The emerging feature of a particularly high $\xi$ from $\nicefrac{B}{J}=-2$ to $\nicefrac{B}{J}\to\infty$ (\textit{zero-derivative curve}) is discussed in the main text. As described in the main text, $F_1(T)\to\infty$ for $B\to 0$, hence leading to a divergence of the $\xi$ ratio. In the plot, all areas with $\xi > 10^{10}$ are shaded in white.}
    \label{fig:heatmap_xi}
\end{figure}

Spin chains constitute an extremely interesting playground for our discussion: already in their simplest version, the nearest-neighbor Ising chain, they exhibit a surprising richness of features. This nearest-neighbor model requires a single parameter to encode the coupling strength; we will call it $J$ instead of $J_1$, for the sake of conciseness. Given the decomposition \ref{Cramer_Rao_Markov_Fisher}, for $\MarkOrd=1$ all the joint probabilities involving more than two spins are fully determined by the one-site and two-sites probabilities. Hence, it makes sense to discuss the relation between the 2-sites joint Fisher information $F_{1:2}(T)$ and the 1-site marginal Fisher information $F_1(T)$. In the same spirit of the Gaussian example in Section \ref{sect:Fisher_in_presence_of_correlations}, we can discuss the inequality
\begin{equation}
    F_{1:2}(T) \overset{?}{\lessgtr} 2 F_1(T).
\end{equation}
Or, what is equivalent, $F_{2|1} \overset{?}{\lessgtr} F_1$ (since $F_{1:2} = F_1 + F_{2|1}$ in this case).
Let us define the ratio
\begin{equation}
    \xi = \frac{F_{1:2}(T)}{2 F_{1}(T)}.
\end{equation}
Super-additive behaviour corresponds to $\xi>1$, and sub-additive  to $\xi<1$; independent random variables yield a perfectly additive behaviour with $\xi = 1$. The value of $\xi$ for a wide choice of parameters is represented in Figure \ref{fig:heatmap_xi}.

It is apparent that there is a significant difference between the behaviours in the ferromagnetic and the anti-ferromagnetic regimes: the first gives $\xi \lesssim 1$ almost everywhere, corresponding to a slightly sub-additive Fisher information. In contrast, the anti-ferromagnetic regime sees significantly high $\xi$ values, representing a strongly super-additive Fisher information. 

\begin{figure*}[htbp]
    \centering
    \includegraphics[width=\textwidth]{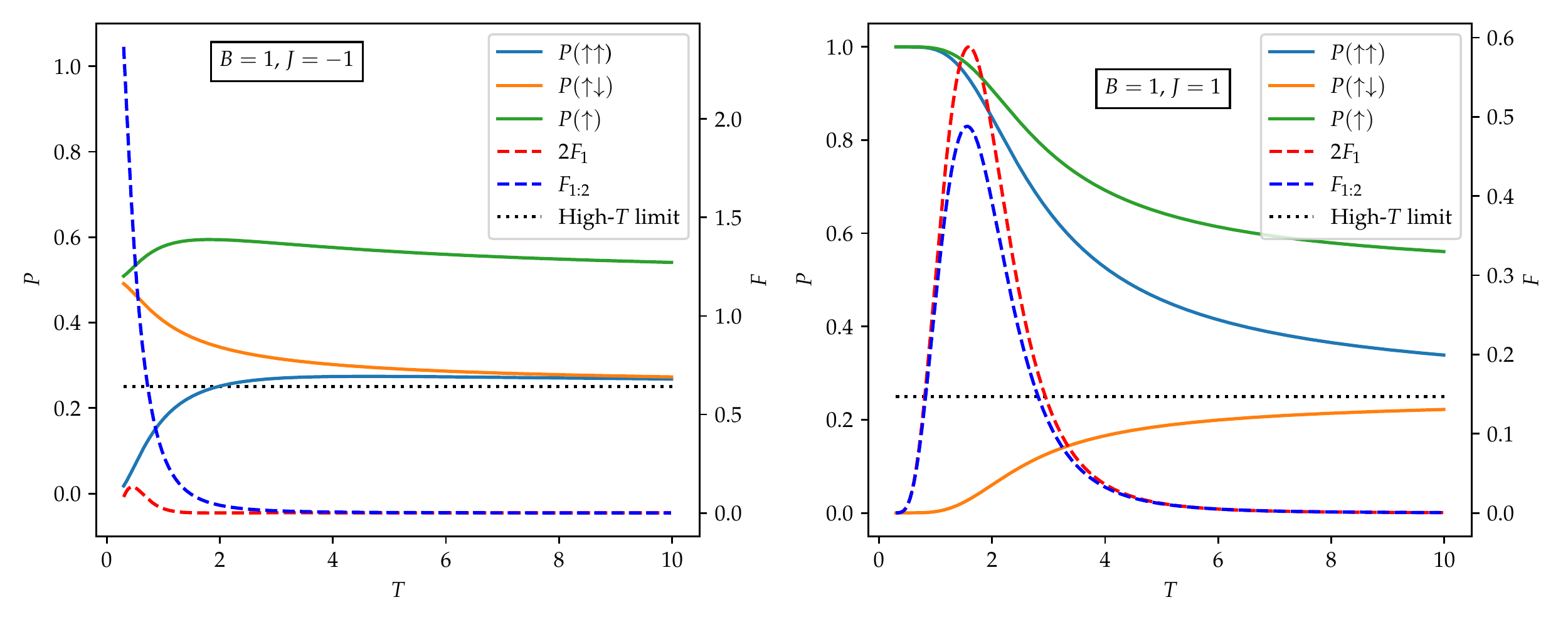}
    \caption{Relevant joint and marginal probabilities on the nearest-neighbors Ising spin chain, represented as a function of the temperature $T$, along with the one-site and two-sites Fisher information.}
    \label{fig:probabilities_Ising_chain}
\end{figure*}

Considering the shape of the different probabilities as a function of the temperature allows us to gain a precious intuition about the conditions under which each regime is attained, as illustrated in  Figure \ref{fig:probabilities_Ising_chain}. Note how, in the $T\to 0$ limit, each curve reaches the probability of the specified configuration in the ground state of the chain: total anti-alignment in the anti-ferromagnetic case ($P(\uparrow\downarrow) = P(\downarrow\uparrow)=\nicefrac{1}{2}$) and total alignment in the ferromagnetic case ($P(\uparrow\uparrow)=P(\uparrow) = 1 )$). 

The $\xi>1$ condition in the anti-ferromagnetic case is mainly due to the strong suppression of the denominator $2F_1(T)$. The reason of the suppression is apparent if one considers how two two-sites joint probabilities add up to give a one-site marginal: the $P(\up\up)$ and $P(\up\down)$ probabilities in the anti-ferromagnetic case exhibit almost opposite derivatives with respect to $T$; hence, their sum $P(\up)$ will show a very mild dependence on the temperature, yielding a reduced Fisher information. This does not happen in the ferromagnetic regime, where the derivatives of $P(\up\up)$ and $P(\up\down)$ have significantly different absolute values. 

This difference at the level of the derivatives can be easily thought in terms of the structure of the ground state of the chain, or, equivalently, of the thermal state for $T\to 0$. In the anti-ferromagnetic case, the ground state is two-fold degenerate (for values of $B$ that are not too large): $P(\up\down)$ and $P(\down\up)$ both have $\nicefrac{1}{2}$ probability, while $P(\up\up) = P(\down\down)=0$. In the high-$T$ limit, all the two-site probabilities will achieve a value of $\nicefrac{1}{4}$. Hence, in the anti-ferromagnetic case $P(\up\down)$ has to lose $\nicefrac{1}{4}$ probability between $T\to 0$ and $T\to\infty$; in the same temperature interval, $P(\up\up)$ has to gain $\nicefrac{1}{4}$; the two derivatives can almost cancel one another.

On the opposite side, the ferromagnetic chain has a non-degenerate ground state: $P(\up\up)$ has to lose $\nicefrac{3}{4}$ of probability in the $T$-interval, while $P(\up\down)$ has only to gain $\nicefrac{1}{4}$: the probabilities cannot exhibit opposite derivatives.

This perspective allows us also to interpret the emerging curve labeled as \textit{zero-derivative curve} in Figure \ref{fig:heatmap_xi} in the anti-ferromagnetic region. The curve corresponds to point of maximum or minimum of the one-site probability $P(\up)$ with respect to the temperature; consequently, they have zero derivative and zero one-site thermal Fisher information, hence causing a divergence of the $\xi$ ratio. Notice that the zero-derivative curve is only visible for a $\nicefrac{B}{J}$ ratio larger than -2. This can be traced back to a fundamental regime change in the structure of the ground state, happening exactly at $\nicefrac{B}{J} = -2$: the anti-ferromagnetic ground state $\ldots\up\down\up\down\up\down\ldots$ gives way to a ferromagnetic ground state $\ldots\up\up\up\up\up\ldots$, due to the frustration of the anti-ferromagnetic interaction operated by the extremely strong external magnetic field. The one-site probability in this high-$B$ regime is monotonous, thus leading to the absence of zero-derivative points. 

Finally, the $\xi$ ratio diverges at all temperatures for $B\to 0$; this is due to $F_1(T) \to 0$ in that limit. In fact, in the absence of an external magnetic field, it does not make any energetic difference for a single spin to be in state $\up$ or $\down$; no information about temperature can therefore be obtained from observations of single spins ($F_1(T)=0$), whereas correlations can still give place to meaningful estimations of temperature ($F_{1:2}(T)\neq 0$). 

We also considered how the introduction of a second-order coupling ($J_2$ in the language of Eq.~\eqref{Hamiltonian_spin_chain}) affects the temperature estimation task. What appears is a much richer phenomenology, summarized in Appendix~\ref{sect:NNN_spin_chain}. The key points discussed above are preserved also in this more general case: chains with predominantly anti-ferromagnetic behaviour tend to give a large advantage, exploiting correlations for temperature estimation ($\xi \gg 1$), whereas chains dominated by ferromagnetic interactions tend to yield a disadvantage~($\xi < 1$). 

\section{Discussion and outlook}
In this paper we studied the Fisher information of correlated stochastic processes. The central result is the decomposition of the joint Fisher information \eqref{decomposition_Fisher_Markov}. It entails that the precision that may be achieved for parameter estimation from any process with finite Markov order obeys an asymptotic bound that scales linearly with the number of outcomes \eqref{Cramer_Rao_Markov_Fisher}.
A finite Markov order does not mean that the correlation is finite; on the contrary, even for $\MarkOrd=1$, all outcomes are correlated, except that these correlations decay exponentially. 
Thus, our results encompass an extremely large class of physical processes (a notable exception being critical systems, for which the correlation length diverges). 
Second, a  central realization of our framework was that, in contrast to entropic information measures, the  Fisher information does not necessarily obey subadditivity, and can be both super or sub-additive, depending on the problem.  
This means that correlations can sometimes help, or sometimes hamper the estimation process, in the sense that the convergence of the mean-squared error with $N$ may be quicker or slower. 
As our results illustrate, generally speaking, anti-correlated outcomes tend to be advantageous, whereas correlated outcomes tend to result in reduced scaling. 

Our focus here was on classical stochastic processes. It is natural to ask about a possible generalization of our main result, Eq.~\eqref{decomposition_Fisher_Markov}, to the quantum domain. 
In this regard, it is useful to distinguish between two scenarios. 
The first is that of continuous quantum measurements~\cite{Wiseman_2009}. In this case, the process is quantum, but the outcomes are classical and local. Indeed, quantum continuous measurements fall \emph{precisely} within the paradigm of our framework, which is thus directly applicable.
For instance, Ref.~\cite{Kiilerich_2014} studied the estimation of the Rabi frequency from photon counting experiments. Their results fall precisely within our framework with $\MarkOrd=1$. 
The same is true for Ref.~\cite{Kiilerich_2016}, which studied homodyne detection.

A classical time series may be also be obtained from a quantum collision model~\cite{Taranto2019,Landi2022,Ciccarello2022}. For instance, we may consider a memory system which is inaccessible to direct observation. At each time step, a fresh probe systems is made to interact with the memory system at and then subject to a fixed, generalised, quantum measurement. Even for Markovian dynamics on the composite space, the resulting time series is in general non-Markovian, with any Markov order realisable. Our results apply directly to this scenario of indirectly probing a quantum system.

More generally, however, one may also consider scenarios requiring non-local measurements, which is often the case for systems exhibiting entanglement or other forms of quantum correlations. 
The generalization of the notion of Markov order to this case has proved to be rather cumbersome \cite{Taranto2018, Taranto2019}: when a completely general measurement is taken into account, it can be shown \cite{Taranto2018} that the quantum Markov order is either 0, 1 or infinite (one could instead try to apply the concept of approximate Markov order~\cite{Fawzi_2015,Kato_2019}).
In fact, it has been shown that the fundamental decomposition~\eqref{decomposition_Fisher}  generally does not hold for the Quantum Fisher information \cite{Micadei2015}; this fact is at the basis of the possibility of achieving higher precision by exploiting specifically quantum properties, such as entanglement.
Furthermore,  when dealing with multi-parameter estimation, one has to take into account the possible non-commutativity of the corresponding operators, thus leading to the necessity of more refined theoretical tools than the Fisher information matrix \cite{Demkowicz2020}. 

To give a simple example of a non-local measurement scenario, suppose one wished to do thermometry, exactly as in Sec.~\ref{sect:application_thermometry_classical_spin_chains}, but considering instead the transverse field Ising model; i.e., where the field is now applied in the $x$ direction.
There has been considerable interest in thermometry of quantum systems (for reviews, see~\cite{Mehboudi_2019,De_Pasquale_2018}).
And, as already discussed in Sec.~\ref{sect:application_thermometry_classical_spin_chains}, in this case the optimal approach is to measure the system in the energy eigenbasis. 
In general, however, this basis is non-local and hence our main result, Eq.~\eqref{decomposition_Fisher_Markov} is not necessarily applicable. 

An extension to the fully-quantum case would be an interesting future avenue of research.

\section*{Acknowledgments}
The authors thank Alessio Benavoli for helpful discussions on the Bayesian approach to parameter estimation. This research was supported by grant number FQXi-RFP-IPW-1910 from the Foundational Questions Institute and Fetzer Franklin Fund, a donor advised fund of Silicon Valley Community Foundation. 
GTL acknowledges the financial support of the S\~ao Paulo Funding Agency FAPESP (Grant No.~2019/14072-0.), and the Brazilian funding agency CNPq (Grant No. INCT-IQ 246569/2014-0). 
MR acknowledges funding by the Irish Research Council under Government of Ireland Postgraduate Scheme grant number GOIPG/2022/2321.

\appendix

\section{Information-theoretic and probability-theoretic definitions of the Markov order}
\label{sect:different_definitions_Markov_order}
In this appendix, we briefly recall the connection between the probability-theoretic and the information-theoretic definition of the Markov order \cite{Crutchfield2003}. 
In \eqref{Markov_order}, we defined the Markov order as
\begin{equation}
    \MarkOrd = \min\left\{\ell \mbox{ s.t. } P(X_0\vert X_{-\ell:0} = P(X_0 \vert X_{-\infty:0}) \right\}.
\end{equation}
Let us now consider the $N$-site joint Shannon entropy $H(X_{1:N})$, for which a well-known chain rule holds \cite{Cover1999}:
\begin{equation}
    H(X_{1:N}) = \sum_{i=1}^N H(X_i \vert X_{i-1}\ldots X_1).
\end{equation}
Each conditional Shannon entropy can be written as:
\begin{equation}
\begin{split}
    &H(X_i\vert X_{i-1}\ldots X_1)\\&= -\sum_{x_1\ldots x_i} P(x_1\ldots x_i) \log P(x_i \vert x_{i-1}\ldots x_1) \\
    & = - \sum_{x_{i-\MarkOrd}\ldots x_i} P(x_{i-\MarkOrd}\ldots x_i) \log P(x_i \vert x_{i-1}\ldots x_{i-\MarkOrd}),
\end{split}
\end{equation}
where we used the probabilistic definition of the Markov order \eqref{Markov_order} in the last equality. Hence, the joint Fisher information can be reduced to:
\begin{equation}
\begin{split}
     H(X_{1:N}) & = - \sum_{i=1}^N \sum_{x_{i-\MarkOrd}\ldots x_i} P(x_{i-\MarkOrd}\ldots x_i)\cdot \\
     & \cdot \log P(x_i \vert x_{i-1}\ldots x_{i-\MarkOrd})  \\
     & = H(X_{1:\MarkOrd-1}) + (N-\MarkOrd) H(X_{1:\MarkOrd}),
\end{split}
\end{equation}
where in the last step the stationarity of the process has been exploited. With the definitions
\begin{equation}
\begin{split}
    & h = H(X_{1:\MarkOrd})\text{, and} \\
    & E = H(X_{1:\MarkOrd-1}) - \MarkOrd H(X_{1:\MarkOrd}),
\end{split}
\end{equation}
we obtain the structure of the information-theoretic definition \eqref{entropy_theoretic_Markov_order}. 

\section{Decomposition of the joint Fisher information}
\label{sect:proof_decomposition_Fisher}
The proof of the decomposition of the joint Fisher information in this appendix is analogous to that in Ref.~\cite{Micadei2015}, but generalized to the case of multi-parameter dependence. Let us consider definition \eqref{Fisher_information_matrix} of the Fisher information matrix for two random variables $X_1, X_2$:
\begin{equation}
\begin{split}
    & \left[F_{X_1 X_2} (\theta)\right]_{ij} \\
    & = \sum_{x_1 x_2} P_\theta(x_1, x_2) \left[\partial_i \log P_\theta(x_1, x_2) \partial_j \log P_\theta(x_1, x_2)\right] \\
    & = \sum_{x_1 x_2} \frac{P_\theta(x_1)}{P_\theta(x_2\vert x_1)}\partial_{\theta_i} P_\theta(x_2\vert x_1) \partial_{\theta_j}P_\theta(x_2\vert x_1) + \\
    & + \sum_{x_1 x_2} \partial_{\theta_i} P_\theta(x_2\vert x_1) \partial_j P_\theta(x_1) + \\
    & + \sum_{x_1 x_2} \partial_{\theta_i} P_\theta(x_1) \partial_{\theta_j} P_\theta(x_2\vert x_1) + \\
    & + \sum_{x_1 x_2} \frac{P_\theta(x_2\vert x_1)}{P_\theta(x_1)} \partial_{\theta_j} P_\theta(x_1) \partial_{\theta_j} P_\theta(x_2\vert x_1)
\end{split}
\end{equation}
Now, notice that the second and the third terms are zero. In fact, for the second:
\begin{equation}
\begin{split}
    & \sum_{x_1 x_2} \partial_{\theta_i} P_\theta(x_2\vert x_1) \partial_{\theta_j} P_\theta(x_1) = \\
    & \sum{x_1} \partial_{\theta_j} P_\theta(x_1) \partial_{\theta_i} \sum_{x_2} P_\theta(x_2\vert x_1), 
\end{split}
\end{equation}
but by normalization of probabilities we have that
\begin{equation}
    \sum_{x_2} P_\theta(x_2 \vert x_1) = 1 \implies \partial_{\theta_i} \sum_{x_2} P_\theta(x_2 \vert x_1) = 0.
\end{equation}
In the same way, it is easy to prove that the third term vanishes. Hence, we have:
\begin{equation}
\begin{split}
    & \left[F_{X_1 X_2} (\theta)\right]_{ij} = \\ & \sum_{x_1 x_2} \frac{P_\theta(x_1)}{P_\theta(x_2\vert x_1)}\partial_{\theta_i} P_\theta(x_2\vert x_1) \partial_{\theta_j}P_\theta(x_2\vert x_1) + \\
    & + \sum_{x_1 x_2} \frac{P_\theta(x_2\vert x_1)}{P_\theta(x_1)} \partial_{\theta_j} P_\theta(x_1) \partial_{\theta_j} P_\theta(x_2\vert x_1) \\
    & = \left[F_{X_1}(\theta)\right]_{ij} + \left[F_{X_2\vert X_1}(\theta)\right]_{ij}.
\end{split}
\end{equation}
The multi-partite case can be proven by iteratively following the same procedure. As a direct consequence, in case of i.i.d. multi-parametric estimation, we have simply:
\begin{equation}
    F_{X_1 X_2}(\theta) = F_{X_1}(\theta) + F_{X_2}(\theta).
\end{equation}

\section{Maximum Likelihood Estimation}
\label{sect:MLE}
In this appendix, we present a brief review of Maximum Likelihood Estimation (MLE), following mainly the book by Kay \cite{Kay1993}, and give some indications about its application to parameter estimation on finite-Markov order processes on a finite sample space. 

Let us consider a parameter-dependent probability distribution $P_\theta(X_{1:N})$, and define the function (of $\theta$):
\begin{equation}
    \ell(\theta) = P_\theta(X_{1:N}),
\end{equation}
which is called \textit{likelihood} function. The value of the parameter $\theta$ proposed by the MLE estimator is given by the maximisation of $\ell(\theta)$ over the entire space in which $\theta$ can take values. This corresponds to choosing the value of $\theta$ which makes the current outcomes as \textit{likely} as possible. 

Notice that the maximising $\theta$ does not change if a monotonous function is applied to $\ell(\theta)$; it is therefore common to maximise the \textit{log-likelihood} $L(\theta)=\log \ell(\theta)$. 

When dealing with processes with finite Markov order $\MarkOrd$ ($\MarkOrd\ll N$) and a finite sample space, the computation of the MLE is greatly eased by making use of the following observations. First, exploiting the definition of the Markov order \eqref{Markov_order}, we can rewrite the log-likelihood as:
\begin{equation}
\begin{split}
    L(\theta) = & \sum_{k=\MarkOrd+1}^N \log P_\theta(X_k \vert X_{k-\MarkOrd:k-1}) + \\
    & + \log P_\theta(X_\MarkOrd\vert X_{1:\MarkOrd-1}) + \ldots \\
    & \ldots + \log P_\theta(X_2 \vert X_1) + \log P_\theta(X_1).
\end{split}
\end{equation}
Notice that all the terms in the summation have the same form: the probability of a certain outcome in the sequence, conditioned on exactly the previous $\MarkOrd$. The remaining terms, conditioned on a smaller number of random variables, give a negligible contribution in the $N\gg \MarkOrd$ limit. Let us now group together the items in the sequence in groups of homogeneous size $\MarkOrd+1$, with a sliding window approach:
\begin{equation}
\begin{split}
    & \eta_1=(X_1, X_2, \ldots, X_\MarkOrd, X_{\MarkOrd+1}); \\
    & \eta_2 = (X_2, X_3, \ldots, X_{\MarkOrd+1}, X_{\MarkOrd+2}). \ldots
\end{split}
\end{equation}
Each of the $\eta$ sequences is associated to a conditional probability in the form:
\begin{equation}
    C_{\eta_j} = P_\theta(X_{j+\MarkOrd} \vert X_{j:j+\MarkOrd-1}).
\end{equation}
Notice that, if any of the variables $X_j$ can take up to $d$ different values, then there are only $z=(\MarkOrd+1)^d$ possible different groups $\eta_j$; the summation can therefore be transferred to the different blocks:
\begin{equation}
    L(\theta) = \sum_{q=1}^z f_q \log C(q) + \mathcal{R},
\end{equation}
where $f_q$ denotes the number of $\eta$ groups having the same form, indexed by $q$, in the sequence, $C(q)$ is the associated conditional probability, and $\mathcal{R}$ denotes the negligible remaining terms. In the common situation $(\MarkOrd+1)^d\ll N$, this allows us to compute the likelihood with a significantly enhanced efficiency with respect to a direct attempt.

We recall also some relevant properties of the MLE. Let us first consider a set $X_{1:N}$ of i.i.d. random variables, extracted from the same $\theta$-dependent distribution $P_\theta(X)$. Under mild regularity conditions on $P_\theta$, it can be proven that \cite{VanDerVaart2000}:
\begin{itemize}
    \item the MLE estimation is \textit{consistent}, meaning that, in the limit $N \to \infty$, the MLE estimator $\hat{\theta}$ is such that $\hat{\theta}\to \theta^*$, where $\theta^*$ is the \textit{true} value of the parameter;
    \item the MLE estimator is \textit{asymptotically normal}, meaning that the random variable $\hat{\theta}-\theta^*$ (where randomness is given by the extraction of different datasets) tends to be distributed according to a Gaussian distribution, with variance given by the Cramér-Rao bound in the form \eqref{Cramér_Rao_iid}. 
\end{itemize}

These propositions were originally formulated for i.i.d. random variables. For the purposes of this paper, it is however fundamental to consider non-i.i.d. random variables. It has been proven \cite{Billingsley_1961} that, under appropriate regularity conditions, they can be extended to Markov chains, once the marginal Fisher information $F_1(\theta)$ is replaced by the conditional Fisher information $F_{2\vert 1}(\theta)$. Furthermore, a process with a finite sample space and of finite Markov order $\MarkOrd$ can be directly translated into a Markov order 1 process by putting together groups of $\MarkOrd$ random variables, in a similar fashion as it is done in the transfer matrix approach, see Appendix \ref{sect:marginalisation_spin_chains}. This allows the extension of the result to larger Markov orders. 

\section{On the meaning of sub-additive Fisher information}
\label{sect:free_lunch}
When dealing with a process yielding a sub-additive Fisher information, it could seem that, by using some estimator able to ignore the correlations, the experimenter could be able to achieve a precision beyond the Cramér-Rao bound. We illustrate the situation with an example.

\begin{figure}[h!tbp]
    \centering
    \includegraphics[width=.9\columnwidth]{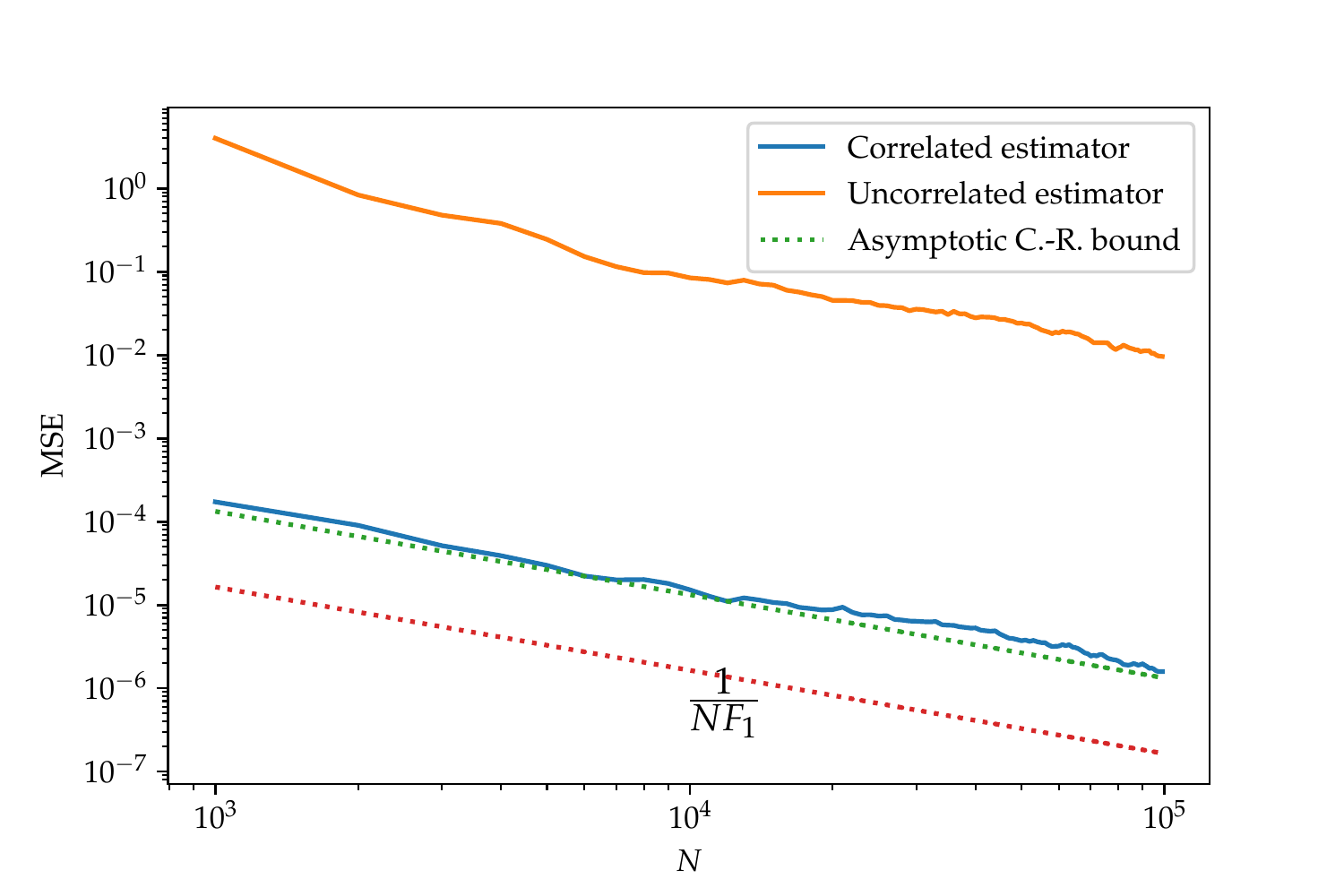}
    \caption{An estimation task for both the regular and the uncorrelated MLE, on the process defined by the transition matrix \eqref{subadditive_example}; the MSE is computed by averaging over 50 repetitions.}
    \label{fig:no_free_lunch}
\end{figure}

Consider the Markov process defined by the transition matrix \eqref{subadditive_example}; we know that it yields a sub-additive Fisher information for $\theta$. Being  a $\MarkOrd=1$ process, its genuine MLE has to take into account the correlations; the likelihood function for a sequence of outcomes $x_{0:N}$ has the form
\begin{equation}
\begin{split}
    \ell(\theta) = & P_\theta(x_N \vert x_{N-1})P_\theta(x_{N-1}\vert x_{N-2})\ldots \\
    & \ldots P_\theta(x_1\vert x_0) P_\theta(x_0).
\end{split}
\end{equation}
We could now define an \textit{uncorrelated} MLE, taking into account only the marginal probabilities, as:
\begin{equation}
    \ell'(\theta) = P_\theta(x_N)P_\theta(x_{N-1})\ldots P_\theta(x_1) P_\theta(x_0).
\end{equation}
Naively, one could think that this uncorrelated MLE should result in a higher precision, with an MSE converging to the i.i.d. version of the Cramér-Rao bound \eqref{Cramér_Rao_iid} instead of the correlated version \eqref{Cramer_Rao_Markov_Fisher}. However, this is not verified, as shown in Figure Figure~\ref{fig:no_free_lunch}. 

This gives an interesting perspective on the meaning of the Fisher information and of the Cramér-Rao bound: they are related to the nature of the \textit{process}, not to the way in which the estimator is designed. In other terms, there is no way of taking advantage of the sub-additivity of the Fisher information to obtain a precision beyond the Cramér-Rao bound.

\section{The Central Limit Theorem for correlated random variables}
\label{sect:CLT}
The usual formulations of the Central Limit Theorem \cite{Billingsley1995} consider i.i.d. variables. Many different possible generalizations of the theorem, relaxing the independence hypothesis have been presented; in this appendix we discuss one that is directly applicable to most cases of physical interest.

Let $X_1, X_2, \ldots$ be a stationary stochastic process, and let us define the coefficients $\alpha_n$ by the inequality
\begin{equation}
    \vert P(X_0 \cap X_n) - P(X_0)P(X_n) \vert \leq \alpha_n.
\end{equation}
In most cases, we will have that $\lim_{n\to\infty} \alpha_n =0$, meaning that two random variables in the stochastic process become increasingly independent for an increasing distance between them. If $\alpha_n \to 0$, the process is called $\alpha$-mixing. It is possible to prove for instance that every Markov process is $\alpha$-mixing, and that $\alpha_n$ decays exponentially on it. 

Let us now consider an $\alpha$-mixing process such that:
\begin{itemize}
    \item $\alpha_n = \mathcal{O}(n^{-5})$
    \item the expectation value $\mathbb{E}[X_0^{12}]$ is finite (hence, due to stationarity, it is finite for all random variables composing the process);
\end{itemize}
then the results of the Central Limit Theorem can be recovered \cite{Billingsley1995}.

Notice that the first hypothesis is easily fulfilled by any process exhibiting exponential decay of the $\alpha_n$ coefficients, such as in the case of Markov processes; the second one is trivially true for any process on a finite space, and also for processes on continuous sample spaces with an exponential depression of the distribution tails, such as normal distributions. 

\section{A Gaussian Markov process}
\label{sect:Gaussian_Markov_process}

In order to verify our claims about the role of the correlation sign in the scaling of the Fisher information, we considered in Fig.~\ref{fig:MSE_Gaussian} a Gaussian process with $\MarkOrd=1$ (i.e., a Gaussian Markov chain). In this section, we will detail its construction. Furthermore, we will use it to illustrate the lack of a relation between the conditional independence required by the definition of Markov order and the independence as defined from the covariances.

A stationary Markov process is fully determined by the joint probability $P(X_1 X_2)$, in terms of which all the relevant conditional and marginal probabilities can be written. We assume this two-sites joint distribution being Gaussian with the form:
\begin{equation}
    P(x_1, x_2) \sim \exp\left[-\frac{1}{2}(\vec{x}-\vec{\mu})^T \sigma^{-1} (\vec{x}-\vec{\mu})\right],
\end{equation}
where $\vec{x} = (x_1, x_2)^T$, $\vec{\mu} = (\mu, \mu)^T$, and the normalization factors have been dropped. Here $\sigma$ is the two-sites covariance matrix, written in terms of the two parameters $\gamma_0 \in \mathbb{R}_+$ and $\rho \in [-1,1]$ as
\begin{equation}
    \sigma = \gamma_0 \begin{pmatrix}1 & \rho \\ \rho & 1\end{pmatrix}.
\end{equation}
Notice that the sign of the nearest-neighbor correlations is encoded in the sign of $\rho$. It is necessary to determine the conditional probability distribution $P(X_2 \vert X_1)$, which we want to sample for the simulation. Like the joint probability, it is also Gaussian, with mean and variance given respectively by
\begin{equation}
\begin{split}
    & E[X_2 \vert X_1] = \mu + \rho(X_1-\mu) \\
    & \Var[X_2 \vert X_1] = \gamma_0 - \rho^2 \gamma_0.
\end{split}
\end{equation}
Using these two relations, we are able to iteratively simulate a Gaussian Markov process.

The process built in this way clearly has Markov order 1 by construction, because only the most recent element from the past plays a role in the probability of the future. The full covariance matrix for this process turns out to be
\begin{equation}\label{Gaussian_sigma_correct}
    \sigma = \gamma_0 \begin{pmatrix} 1 & \rho & \rho^2 & \rho^3 & \ldots \\ \rho & 1 & \rho & \rho^2 & \ldots \\ \rho^2 & \rho & 1 & \rho & \ldots \\ \vdots & \vdots & \vdots & \vdots & \ddots \end{pmatrix}.
\end{equation}
It is rather instructive to compare this behaviour with another possible definition of a Gaussian process; let us consider the Gaussian joint probability distribution with mean vector $\vec{\mu} = (\mu, \mu, \ldots, \mu)$ and covariance matrix
\begin{equation}\label{Gaussian_sigma_naive}
    \sigma = \gamma_0 \begin{pmatrix}1 & \rho & 0 & 0 & \ldots \\ \rho & 1 & \rho & 0 & \ldots \\ 0 & \rho & 1 & \rho & \ldots\\ \vdots & \vdots & \vdots & \vdots & \ddots  \end{pmatrix}.
\end{equation}
In this case, the covariances for a distance larger than 1 are exactly zero; however, computing explicitly the conditional probabilities, we find that a probability of the form $P(X_3 \vert X_2 X_1)$ depends non-trivially on $X_1$, hence it is necessarily different from $P(X_3 \vert X_2)$: despite its appearance, 
Eq.~\eqref{Gaussian_sigma_naive} does not have Markov order 1. Instead, its Markov order is infinite. 
The process with actual Markov order 1 is that given by Eq.~\eqref{Gaussian_sigma_correct}.

\section{Marginalisation of spin chains}
\label{sect:marginalisation_spin_chains}
A detailed description of the method to obtain the marginals has been provided in~\cite{Feldman1998}; here we only give a brief review, considering for the sake of simplicity a $R=1$ spin chain with local dimension $d=2$ (a spin-$\nicefrac{1}{2}$ Ising chain).

The Gibbs state \eqref{thermal_state_spin_chain} of $N$ spins can be conveniently rewritten in terms of an iterated product of components of the so-called \textit{transfer matrix} $V$ (note that periodic boundary conditions are assumed):
\begin{equation}
    p\left(s_1,\ldots,s_N\right) = \frac{1}{Z} V(s_1,s_2)V(s_2,s_3)\ldots V(s_N,s_1)
    \label{joint_transfer_matrix}
\end{equation}
where the transfer matrix is defined as \cite{Yeomans1992}:
\begin{equation}
    V = \begin{bmatrix}\exp\left[(J_1+B)/T\right] & \exp\left[-J_1/T\right] \\ \exp\left[-J_1/T\right] & \exp\left[(J_1-B)/T\right] \end{bmatrix},
\end{equation}
and it is well-known that the partition function $Z$ of $N$ spins can be written as:
\begin{equation}
    Z = \Tr\left[V^N\right].
    \label{partition_function}
\end{equation}

We would now marginalize $p(s_1,\ldots,s_N)$ over all the spins but the first $m$, i.e., to compute:
\begin{equation}
    p(s_1,\ldots,s_m) = \sum_{s_{m+1},\ldots,s_N} p(s_1,\ldots,s_N).
\end{equation}
Note that, since the periodic boundary conditions grant translational invariance, the obtained expression is the general form of an $m$-spin marginal, and not specific to the first $m$ spins. In terms of the transfer matrix expression \eqref{joint_transfer_matrix} we have therefore:
\begin{equation}
    \begin{split}
        p(s_1,\ldots,s_m) = & \frac{1}{Z} V(s_1,s_2) \cdot\ldots\cdot V(s_{m-1},s_m) \\ & \cdot\sum_{\stilde{m+1},\stilde{N}} V(s_m,\stilde{m+1})V(\stilde{m+1},\stilde{m+2})\cdot\\ & \ldots\cdot V(\stilde{N-1},\stilde{N}) V(\stilde{N},s_1).
    \end{split}
\end{equation}
but the summation over the spins $\stilde{m+1},\ldots,\stilde{N}$ is equivalent \cite{Dobson1969} to a matrix multiplication. Thus:
\begin{equation}
    \begin{split}
        p(s_1,\ldots,s_m) &  = \frac{1}{Z} V(s_1,s_2)\cdot\ldots\cdot V(s_{m-1},s_m) \cdot \\ & \cdot \left[V^{N-m+1}\right]_{s_1,s_m} \\ & = \frac{1}{Z}\left[V^{N-m+1}\right]_{s_1,s_m} \prod_{j=1}^{m-1} V(s_j,s_{j+1}).
    \end{split}
\end{equation}
This expression has been obtained without any approximation, and holds for any value of $N$. However, we are interested in the thermodynamic limit $N\to\infty$. It is noteworthy that only the dominant eigenvalue $\lambda$ of the transfer matrix and the corresponding eigenvector $u$ play a role in this limit, yielding the final expression:
\begin{equation}
    p(s_1,\ldots,s_m) = \frac{[u]_{s1}[u]_{s_m}}{\lambda^{m-1}} \prod_{j=1}^{m-1} V(s_j,s_{j+1}).
\end{equation}
 
A procedure to generalize the transfer matrix to higher-$d$ or higher-$R$ chains is detailed in \cite{Dobson1969}, and involves the creation of a $d^R\times d^R$-dimensional matrix, whose components are denoted by multi-indices built on $R$ spins (here we will only put the relevant indices one after the other, like $s_i s_j$). The obtained transfer matrix is in general non-symmetric, and it is hence not possible to guarantee that it can be fully diagonalised. However, it can be proven \cite{Feldman1998,Ruelle1999} that the dominant eigenvalue of $V$ still dominates the trace (and the partition function, according to \eqref{partition_function}); the corresponding left and right eigenvectors ($u^{\mathcal{L}}$ and $u^{\mathcal{R}}$ respectively) have non-negative components. 

It is important here to mention the fact that, unlike what happens in the $R=1$ scenario, when dealing with longer ranges $R$ the transfer matrix only permits the direct computation of marginals of $m=kR$, $k\in\mathbb{N}$ spins. Any marginal consisting of a number of spins which is not a multiple of $R$ has to be computed by manual marginalization of a larger marginal. The general expression for a $m=kR$ marginal of a range-$R$ spin chain is:
\begin{equation}
    \begin{split}
        & P_T(s_1,\ldots,s_{kR}) = \frac{\left[u^{\mathcal{L}}\right]_{s_1\ldots s_R} \left[u^{\mathcal{R}}\right]_{s_{(k-1)R+1}\ldots s_{kR}}}{\lambda^{k-1}} \cdot \\
        & \cdot \prod_{j=0}^{k-2} V\left(s_{jR+1}\ldots s_{(j+1)R} ; s_{(j+1)R+1}\ldots s_{(j+2)R}\right).
    \end{split}
    \label{marginals_spin_chain_general}
\end{equation}

\section{Markov order of spin chains}
\label{sect:Markov_order_spin_chains}
The proof of $\MarkOrd \leq R$ is fully general, and has been first presented in~\cite{Feldman1998b}. Let $R$ be the interaction range, and consider the calculation of a marginal over $L$ spins, where $L=RL'$ for a suitable $L'\in \mathbb{N}$. Let us introduce a multi-index notation to describe the behavior of a whole block of $R$ spins as:
\begin{equation}
    \eta_i = s_{Ri}s_{Ri+1}\ldots s_{R(i+1)-1}.
\end{equation}
The marginal probability distribution for a sequence of length $L$ is then given by:
\begin{equation}
    P(s_0,\ldots,s_{L-1}) = \frac{u^{\mathcal{R}}_{\eta_{L'-1}} u^{\mathcal{L}}_{\eta_0}}{\lambda^{L'-1}} \prod_{i=1}^{L'-2} V(\eta_i,\eta_{i+1}).
\end{equation}

Let us now consider two sequences, both of length $L$, centered in position 0, called $\overleftarrow{s}^L = s_{-L} s_{-L+1}\ldots s_{-1}$ and $\overrightarrow{s}^L=s_0 s_1 \ldots s_{L-1}$. We want to express the conditional probability:
\begin{equation}
    P(\overrightarrow{s}^L\vert\overleftarrow{s}^L) = \frac{P(\overrightarrow{s}^L, \overleftarrow{s}^L)}{P(\overleftarrow{s}^L)}.
\end{equation}
Using the marginal probability detailed before:
\begin{equation}
    P(\overrightarrow{s}^L\vert\overleftarrow{s}^L) = \frac{u^{\mathcal{R}}_{\eta_{L'-1}}}{\lambda^{L'} u^{\mathcal{R}}_{\eta_{-1}}} \prod_{i=-1}^{L'-2} V(\eta_i,\eta_{i+1}),
\end{equation}
from which it is apparent that of all the spin blocks in $\overleftarrow{s}^L$, the conditional probability only depends on the single spin block $\eta_{-1}$, since it is the only one which appears in the expression. One has finally:
\begin{equation}
    P(\overrightarrow{s}^L\vert\overleftarrow{s}^L) = P(\overrightarrow{s}^L\vert \eta_{-1}).
\end{equation}
The thesis follows, because $\eta_{-1}$ has length $R$. 

In this paper, we have only considered nearest-neighbors chains, with $R=1$. It is rather easy to prove for such chains that, for $T\neq 0, \infty$, $\MarkOrd=1$. Notice that the condition $R=1$ implies that $J_1 \neq 0$, by definition of $R$. 

Proceeding by contradiction, let us assume that the Markov order is smaller than the interaction range, hence $\MarkOrd = 0$, $R=1$. Therefore the spins behave independently and the conditionals can be dropped:
\begin{equation}
    P(s_2 \vert s_1) = P(s_1),
\end{equation}
or, equivalently, the probability factorizes, as
\begin{equation}
    P(s_1 s_2) = P(s_1)P(s_2).
    \label{factorization_Ising}
\end{equation}
The expression for both the two-sites joint probabilities $P(s_1 s_2)$ and the one-site marginals are analytically known (but lengthy). Plugging them into \eqref{factorization_Ising} and solving for $J$, the only possible solution we obtain is $J=0$, which contradicts the assumption $R=1$. Hence, for the Ising chain we have, at finite temperature, $\MarkOrd=R$. 

\section{Advantage and disadvantage on NNN spin chains}
\label{sect:NNN_spin_chain}
\begin{figure*}
    \subfloat[$J$=2$B$]{\includegraphics[width=.4\textwidth]{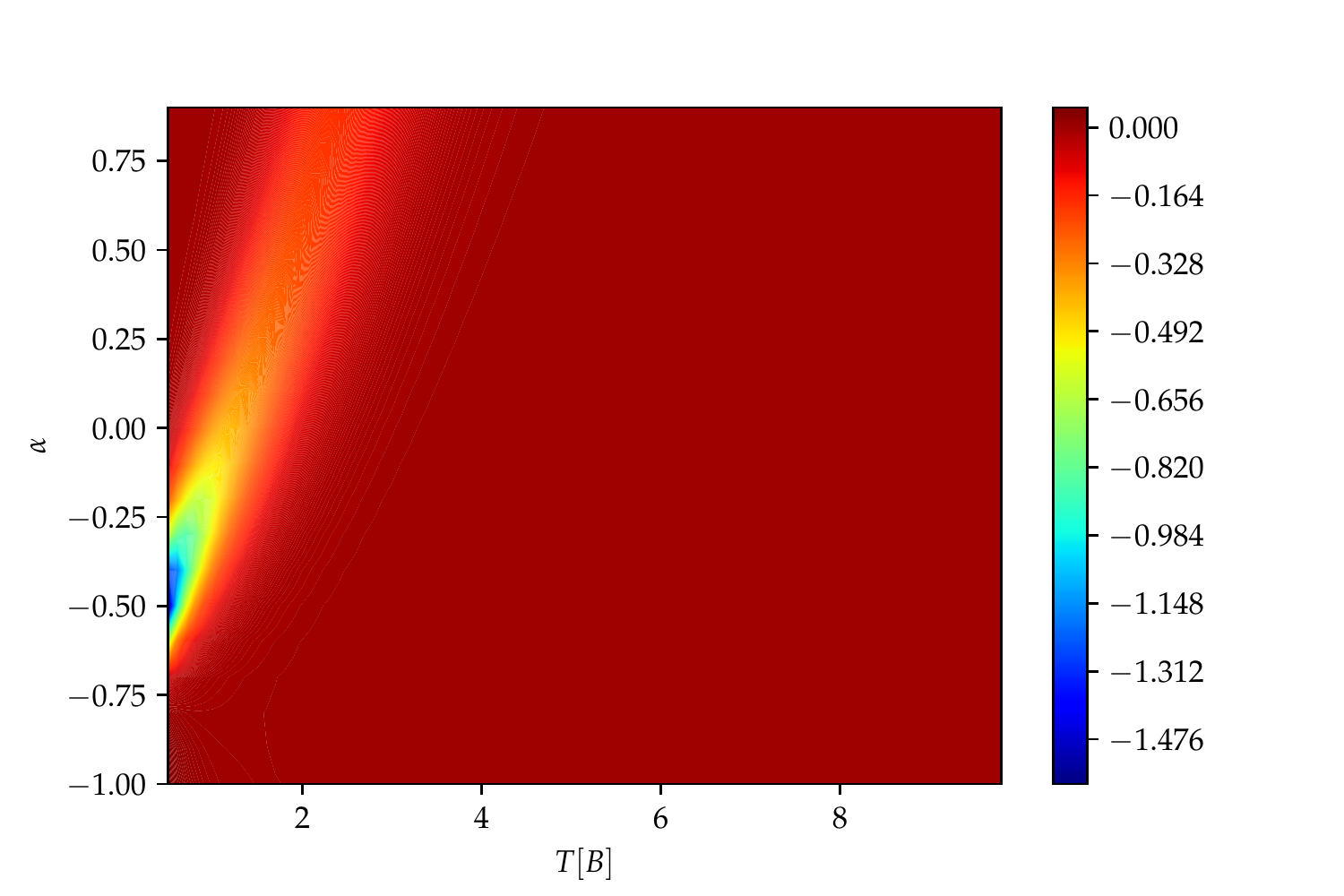}}
    \subfloat[$J$=-2$B$]{\includegraphics[width=.4\textwidth]{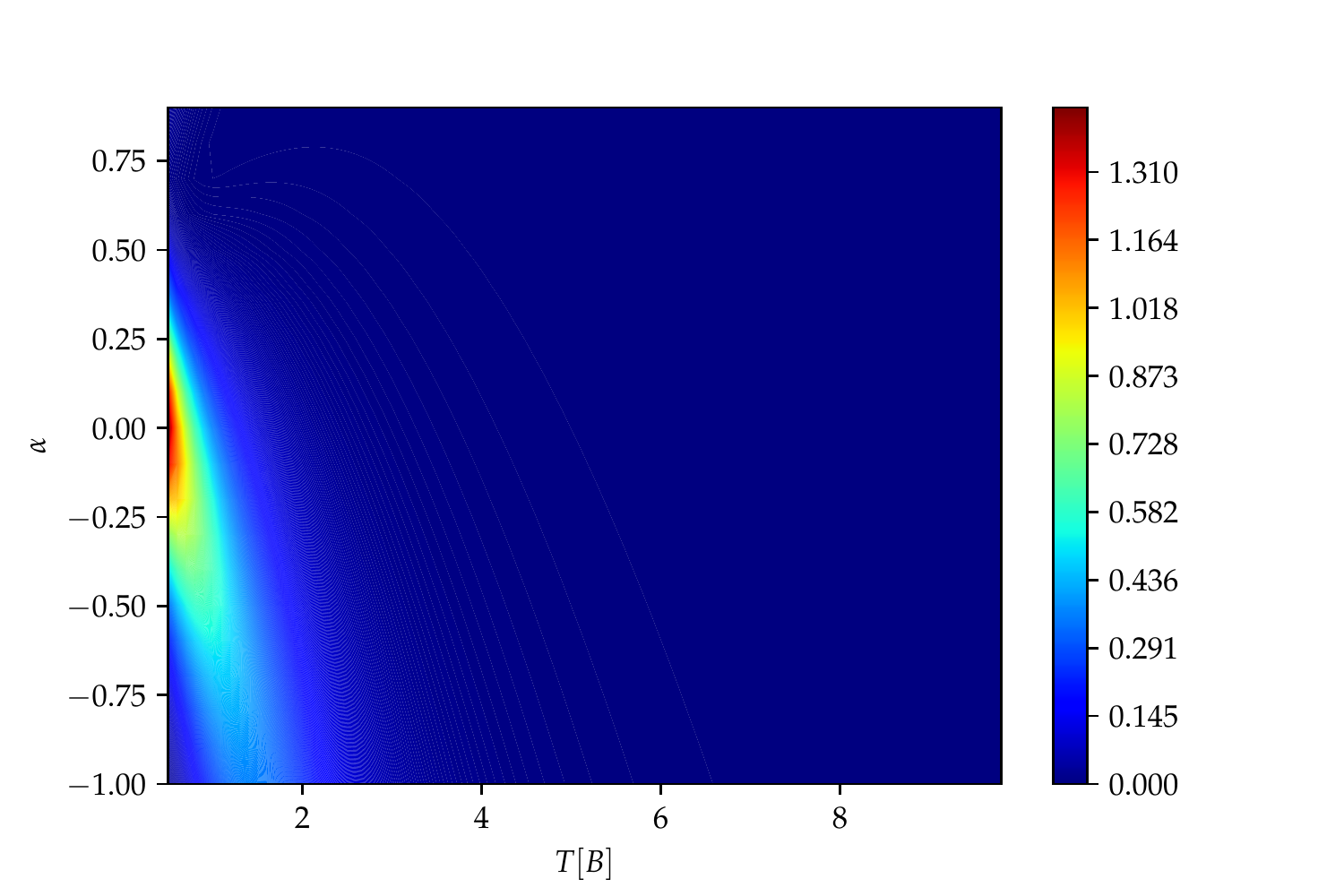}}\\
    \subfloat[$J$=0.1$B$]{\includegraphics[width=.4\textwidth]{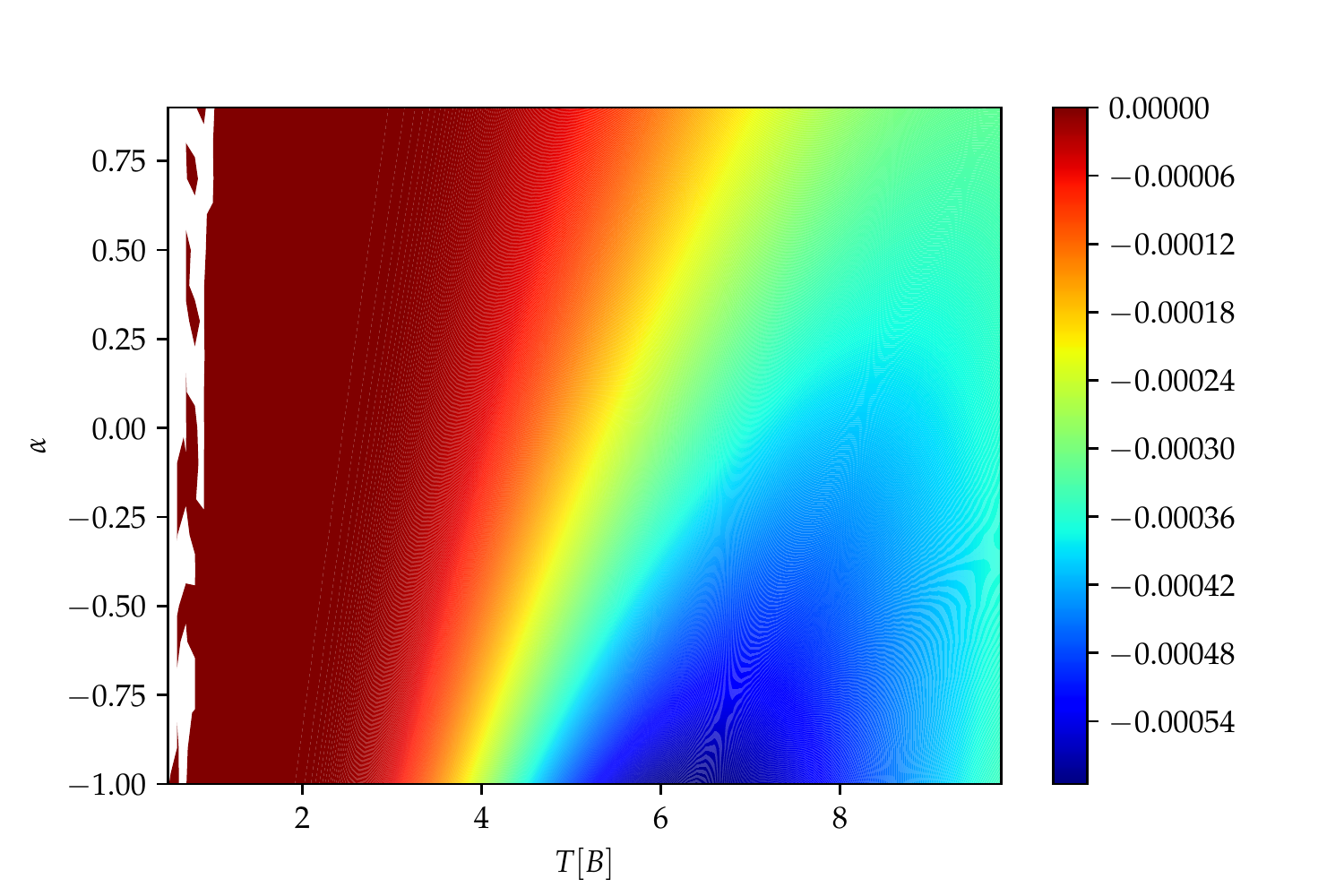}}
    \subfloat[$J$=-0.1$B$]{\includegraphics[width=.4\textwidth]{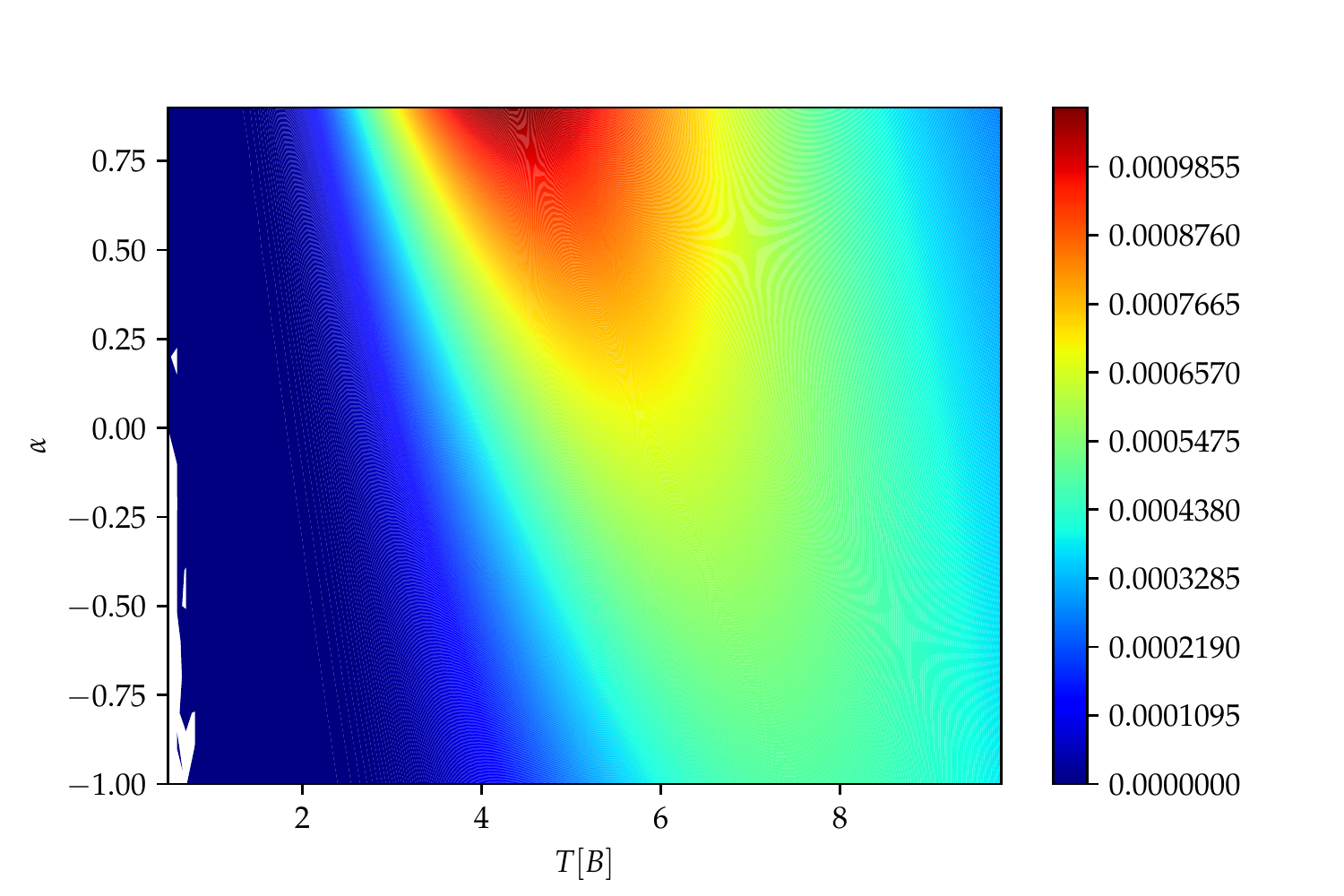}}
    \caption{Behaviour of $\Delta F$ for next-to-nearest-neighbour for different values of the first-order coupling parameter $J$, as a function of temperature $T$ and relative coupling strength $\alpha$. Negative values (disadvantage) occur for all cases of positive correlation (panels~(a) and~(c)); positive values (advantage) occur in all cases with negative correlations (panels~(b) and~(d)).}
    \label{fig:NNN_chain_delta_F}
\end{figure*}

In this appendix, we extend the spin chain thermometry from the main body to a higher coupling order. In particular, we analyze a NNN spin chain with a Hamiltonian of the form
\begin{equation}
    \hat{H} = - B \sum_j \hat{\sigma}_j^z - J \sum_j \hat{\sigma}_j^z \hat{\sigma}_{j+1}^z - \alpha J \sum_j \hat{\sigma}_j^z \hat{\sigma}_{j+2}^z,
\end{equation}
where $\alpha$ represents the relative strength between the first and the second-order couplings. In this case, it is more instructive to discuss the advantage in terms of a difference instead of a ratio: let $\Delta F = F_{1:2} - 2F_1$. This choice is due to the fact that, for some of the plots represented, the value of $F_1$ is so small that the ratio $\xi$ would become difficult to represent, even on a logarithmic scale. Metrological advantage corresponds to $\Delta F > 0$, disadvantage to~$\Delta F < 0$. Figure~\ref{fig:NNN_chain_delta_F} shows the behaviour of $\Delta F$ for different regimes of $J$ and $B$. 

We notice that, even though the NNN spin chain exhibits a richer phenomenology than the NN Ising chain, the relevant points are the ones already mentioned in the previous parts. Again, for $J>0$ (primary ferromagnetic), one has disadvantage, for $J<0$ (primary anti-ferromagnetic) advantage arises. In presence of a very strong external magnetic field, alignment is also enforced in the otherwise anti-ferromagnetic chain; in this case, the behaviour is similar to the ferromagnetic case, with $\Delta F < 0$. Overall, weaker coupling (relative to $B$) leads to weaker $\Delta F$.

\bibliography{bibliography}
\end{document}